\documentclass[preprint,showpacs,superscriptaddress]{revtex4-1}

\usepackage[document]{ragged2e}
\usepackage{bbm}
\usepackage{amsmath}
\usepackage{hyperref}
\usepackage{graphicx}
\usepackage{graphics}
\usepackage{amssymb}
\usepackage{mathtools}
\usepackage{xcolor}
\usepackage{physics}
\usepackage[british]{babel}
\usepackage{csquotes}
\usepackage{graphicx}
\usepackage{lineno}

\begin{document}

\title{Fourier Transform Noise Spectroscopy}

\author{Arian Vezvaee}
\thanks{These authors have contributed equally to this work.}
%\thanks{Current affiliation: Department of Electrical and Computer Engineering, and Center for Quantum Information Science and Technology, University of Southern California, Los Angeles, California, 90089, USA}
\affiliation{Department of Chemistry, University of Colorado Boulder, Colorado 80309, USA}

\author{Nanako Shitara}
\thanks{These authors have contributed equally to this work.}
\affiliation{Department of Chemistry, University of Colorado Boulder, Colorado 80309, USA}
\affiliation{Department of Physics, University of Colorado Boulder, Colorado 80309, USA}

\author{Shuo Sun}
\affiliation{Department of Physics, University of Colorado Boulder, Colorado 80309, USA}
\affiliation{JILA, University of Colorado Boulder, Colorado 80309, USA}

\author{Andr\'{e}s Montoya-Castillo}
\email{Andres.MontoyaCastillo@colorado.edu}
\affiliation{Department of Chemistry, University of Colorado Boulder, Colorado 80309, USA}

%\linenumbers

%%%%%%%%%%%%%%%%%%%%%%%%%
%%%%%%%%%%%%%%%%%%%%%%%%%

\begin{abstract}
Spectral characterization of noise environments that lead to the decoherence of qubits is critical to developing robust quantum technologies. While dynamical decoupling offers one of the most successful approaches to characterize noise spectra, it necessitates applying large sequences of $\pi$ pulses that increase the complexity and cost of the method. Here, we introduce a noise spectroscopy method that utilizes only the Fourier transform of free induction decay or spin echo measurements, thus removing the need for the application many $\pi$ pulses. We show that our method faithfully recovers the correct noise spectra for a variety of different environments (including $1/f$-type noise) and outperforms previous dynamical decoupling schemes while significantly reducing their experimental overhead. We also discuss the experimental feasibility of our proposal and demonstrate its robustness in the presence of statistical measurement error. Our method is applicable to a wide range of quantum platforms and provides a simpler path toward a more accurate spectral characterization of quantum devices, thus offering possibilities for tailored decoherence mitigation.


\end{abstract}

\maketitle

\section*{Introduction}

Nearly all current quantum technology applications rely on a two-level quantum system (qubit) that is subject to environmental noise. In the pure dephasing limit this environmental noise causes fluctuations in the frequency of the qubit that lead to decoherence. Spectral characterization of such environments is the most crucial step in successfully controlling and suppressing decoherence. Indeed, characterizing the noise spectrum allows for a filter-design approach that suppresses the noise and improves the coherence of the qubit~\cite{Cywinski2008PRB,Biercuk2011JP,Uhrig2007PRL,Biercuk2009PRA}. Therefore, developing methods that can recover the noise spectrum of qubit environments has been one of the most active fields of research over the past two decades~\cite{Yang2010Fphys, Szakowski2017JP,DegenRevModPhys2017,Suter2016Review}. Among these efforts, dynamical decoupling noise spectroscopy (DDNS) ~\cite{Alvarez2011PRL,Yuge2011PRL,Norris2016PRL,Krzywda2019NJP} has been one of the most successful approaches. In this method, applying a sequence of $\pi$-pulses turns the qubit into a noise probe (approximated as a frequency comb) that isolates contributions from particular frequencies of the noise spectrum. The dynamical decoupling framework has been studied extensively theoretically and implemented experimentally in various platforms such as superconducting circuits~\cite{Bylander2011NatPhys, SungNature2019}, ultracold atoms~\cite{Almog2011JP}, quantum dots~\cite{Dial2013PRL,farfurnik2023all,Connors2022NatCom}, and nitrogen-vacancy (NV) centers in diamonds~\cite{BarGill2012NatComm,Romach2015PRL}. A DDNS protocol based on the Carr-Purcell-Meiboom-Gill (CPMG) sequence~\cite{Carr1954,Meiboom1958} was proposed by \'Alvarez and Suter~\cite{Alvarez2011PRL} which would ideally yield a system of equations and unknowns from the measured values of the qubit coherence $C(t)= |\left\langle\rho_{01}(t)\right\rangle|/|\left\langle\rho_{01}(0)\right\rangle|$, and specific frequencies of the spectrum. However, this method offers reasonable performance only when the number of $\pi$-pulses in each sequence is large. Beyond a pulse economy standpoint, other difficulties, such as deviations from the ideal frequency comb approximation~\cite{Szankowski2018PRA}, have recently inspired utilizing neural networks as `universal function approximators' to reconstruct the noise spectrum from the coherence function of the qubit~\cite{Wise2021PRX}. The success of this deep learning method suggests the existence of a one-to-one mapping between the two quantities. 

Here, we present a simple and inexpensive method that uniquely maps the measured coherence function of a qubit to its noise power spectrum, removing the need for long sequences of $\pi$-pulses at the heart of DDNS or turning to neural networks. In fact, we show that the map obtained using neural networks in Ref.~\cite{Wise2021PRX} can be found explicitly and analytically and then translated to a simple and effective noise spectroscopy method. This approach only requires free induction decay or spin-echo measurements of the qubit and employs a simple Fourier transform to accurately reconstruct the noise spectrum of the system. While Fourier spectroscopy has been implemented in Nuclear Magnetic Resonance and on different types of quantum processors~\cite{DegenRevModPhys2017,Yan2012PRB,Boss2016PRL}, it has not been utilized in the context of pure dephasing with the filter function formalism. Here, we combine the Fourier transform technique with the filter function formalism to introduce an approach we call Fourier transform noise spectroscopy (FTNS) that significantly enhances one's ability to reconstruct the power spectrum while dramatically reducing the required experimental overhead. We show that FTNS enables the reconstruction of the noise spectrum over a frequency range that is otherwise inaccessible through DDNS --- information that is critical for effective noise mitigation. We then extend the FTNS method to directly extract the noise spectrum from a spin-echo signal, which becomes necessary when the system of interest is dominated by strong low-frequency noise. While this FTNS method requires taking two time derivatives of the signal and is therefore sensitive to measurement noise in the time domain, we show that simple signal processing steps can mitigate the effect of such noise and yield accurate results.

\section*{Results and discussions} \label{sec-DDNS}

\subsection*{\textbf{Theoretical description}}

We begin by laying out the theoretical basis for the filter function formalism in a pure dephasing setup~\cite{Cywinski2008PRB,Yuge2011PRL,Szakowski2017JP,Gu2019}. In this case, the qubit relaxation process (quantified by $T_1$) takes much longer than phase randomization (quantified by $T_2^\star$), implying that the decoherence time $T_2^{-1} = (2T_1)^{-1}+T_2^{\star{-1}} \approx T_2^{\star{-1}}$ becomes a measure of how fast the phase information is lost due to environmental fluctuations. Frequency fluctuations of a qubit subject to a stationary, Gaussian noise, $\hat\beta(t)$, can be described by the Hamiltonian $\hat{H}=\frac{1}{2}[\Omega+\hat\beta(t)] \hat{\sigma}_{z}$, where $\Omega$ is the natural frequency of the qubit. Here, the coherence function is $C(t)=\mathrm{e}^{-\chi(t)}$, where the attenuation function $\chi(t)$ is given by the overlap of the noise spectrum and a filter function that incorporates the effect of the pulses on the system:
\begin{equation} \label{eq-coherence-def}
    \chi(t)= -\text{ln}[C(t)]=\frac{1}{4 \pi} \int_{-\infty}^{\infty} d \omega~ S(\omega)F(\omega t). 
\end{equation} 
The noise spectrum, $S(\omega)=\int_{-\infty}^{\infty}d t~ \mathrm{e}^{\mathrm{i} \omega t} S(t)$, is the Fourier transform of the equilibrium time correlation function of the environmental noise, $S(t)=\langle\{\hat\beta(t),\hat\beta(0)\}\rangle/2$, where $\{A,B\} = AB + BA$ is the anticommutator. The filter function, $F(\omega t)$, encodes the sign switching ($\pm 1$) of the environmental fluctuations upon application of each $\pi$ pulse in the sequence~\cite{Cywinski2008PRB}.

The use of the absolute value in the definition of $C(t) \propto |\left\langle\rho_{01}(t)\right\rangle|$ merits further comment. Without the absolute value, $\tilde{C}(t) = \left\langle\rho_{01}(t)\right\rangle$ contains both a real and an imaginary component, which is the output of the full coherence measurement, i.e., $\langle \sigma_\mathrm{x}(t)\rangle + \mathrm{i}\langle \sigma_\mathrm{y}(t)\rangle$. Here, $\langle \sigma_\mathrm{x}(t)\rangle $ refers to the Ramsey measurement of the real part that involves the sequence \newline $R_\mathrm{Y}(\pi/2)~-~t~-~R_\mathrm{Y}(-\pi/2)$, giving access to Re$[\rho_{01}(t)]$, whereas $\langle \sigma_\mathrm{y}(t)\rangle $ refers to the Ramsey measurement of the imaginary part that involves the sequence $R_\mathrm{Y}(\pi/2)~-~t~-~R_\mathrm{X}(\pi/2)$, giving access to Im$[\rho_{01}(t)]$~\cite{SungNature2019}. For quantum noise sources that obey Gaussian statistics, this measurement can be written as $\tilde{C}(t) \sim \mathrm{e}^{-\chi(t) + \mathrm{i}\Phi(t)}$~\cite{PazSilva2017PRA, Kwiatkowski2020PRB,Mukamel1985,mukamel1995}. We consider the absolute value of this measurement, which leads to $C(t) = |\tilde{C}(t)| \sim \mathrm{e}^{-\chi(t)}$. While removing the dependence on $\Phi(t)$ may appear to cause information loss, it is not so as $\Phi(t)$ contains the same information about the noise spectrum as $\chi(t)$. Indeed, $\Phi(t)$ is related to $\chi(t)$ via detailed balance, with: 
\begin{equation}
 \Phi(t)=  \int_{-\infty}^{\infty} d \omega~  S(\omega)\coth\Big(\frac{\omega}{2k_\mathrm{B}T}\Big) G(\omega t),
\end{equation}
where $G(\omega t)$ encodes the effect of the DD sequence on the imaginary-part Ramsey procedure, $T$ denotes temperature, and $k_\mathrm{B}$ the Boltzmann constant. Hence, knowledge of either $\chi(t)$ or $\Phi(t)$ implies knowledge of the other. Other noise spectroscopy works have distinguished between classical and quantum noise sources, with classical noise leading to a signal where $C(t) \sim \mathrm{e}^{-\chi(t)}$. However, such a measurement would indicate the breakdown of detailed balance. Instead, we articulate the problem in terms of $C(t) = |\tilde{C}(t)|$ and emphasize that such a measurement does not imply that the source of noise is classical. We also note that previous work has shown that $\Phi(t)$ appears in the case of biased coupling~\cite{Kwiatkowski2020PRB} or in the M2 model~\cite{PazSilva2017PRA}, when the interaction of the qubit with the bath has the form $\frac{1}{2} \lambda (\hat{\sigma}_\mathrm{z} + \eta \hat{\mathbb{I}}) \otimes \hat{V}$, where $\hat{V}$ is a bath operator and $\eta \neq 0$. This case is particularly relevant for qubits based on the $m=0,\pm 1$ levels of the NV center in diamond.

To demonstrate the advantages of our proposed FTNS, we first consider what is arguably the state-of-the-art approach to noise spectroscopy: the \'Alvarez-Suter protocol. The main insight of the \'Alvarez-Suter method is that when the number of pulses is sufficiently large, the filter function reaches the spectroscopic limit. In this limit, one can approximate the filter function by a $\delta$-function (frequency comb) with various harmonics: $\chi(t) \approx t \sum_{k=1}^{k_\mathrm{c}}\left|A_{k\omega_0}\right|^{2} S\left(k \omega_{0}\right)$, where $A_{k\omega_0}$ are the Fourier coefficients for a given pulse sequence, truncated at $k_\mathrm{c}$ (for the CPMG sequence, $A_{k\omega_0}=0$ for even $k$). Applying many $\pi$-pulses is necessary for each peak to better resemble a $\delta$-function. The extreme case of $k_\mathrm{c}=1$ approximates the filter function as a single $\delta$ function, discarding many details of the noise spectrum. This is referred to as the single $\delta$-function approximation or the first harmonic approximation. Often, one can still account for a limited number of harmonics (set by the cut-off $k_\mathrm{c}$), which attenuates the loss of spectral information~\cite{Szakowski2017JP,Wise2021PRX}. In the latter case, by appropriately varying the delay time between pulses and the total time of the sequence, one can form a linear system of equations consisting of coherence values at selected times and a matrix of contributing Fourier coefficients. Inverting this system of equations yields the noise spectrum at the probed frequencies, which are bounded by $\pi/\tau_\mathrm{max} \leq|\omega^{\rm DDNS}|\leq \pi/\tau_\mathrm{min}$. Here, $\tau_\mathrm{max (min)}$ is the maximum (minimum) delay between consecutive $\pi$-pulses required to minimize the overlap between subsequent pulses and validate the instantaneous pulse assumption. Furthermore, since $A_{(k=0)}=0$ for balanced pulse sequences like CPMG, the zero-frequency part of the spectrum cannot be accessed directly. Thus, going beyond the $\pi/\tau_\mathrm{max} \leq |\omega^{\rm DDNS}|\leq \pi/\tau_\mathrm{min}$ limit and extracting $S(\omega=0)$ requires \textit{imbalanced} sequences such as concatenated dynamical decoupling (CDD)~\cite{Norris2016PRL}. Hence, the experimental overhead, frequency restrictions, and accuracy dependence on harmonic inclusions of \'Alvarez-Suter~\cite{Szankowski2018PRA} motivate the development of a more accessible scheme.

\subsection*{\textbf{FTNS directly maps FID coherence to the noise spectrum}} \label{sec-map}

We introduce a radically more straightforward approach by inverting Eq.~\eqref{eq-coherence-def} directly to obtain the noise power spectrum. We first demonstrate this in the context of free induction decay, noting that $F_\text{FID}(\omega t)=(4/\omega^{2}) \sin ^{2}(\omega t / 2)$~\cite{Cywinski2008PRB}. Substituting $F_\text{FID}(\omega t)$ in Eq.~\eqref{eq-coherence-def}, and differentiating twice with respect to time, we obtain
\begin{equation} \label{eq-chi-derivative}
     \ddot{\chi}_{\mathrm{FID}}(t)=\frac{1}{2\pi} \int_{-\infty}^{\infty} d \omega\ S(\omega)\cos(\omega t).
\end{equation}
We Fourier transform both sides to find 
\begin{equation} \label{eq-ft-ns}
    S(\omega)=\sqrt{2 \pi }~\mathcal{F}\big[\ddot{\chi}_{\mathrm{FID}}(t)\big],
\end{equation}
noting that $S(-\omega)=S(\omega)$. This straightforward derivation demonstrates that there is a simple and invertible one-to-one map between the noise power spectrum $S(\omega)$ and the second time derivative of the logarithm of the experimentally measured coherence function.

%%%%%%%%%%%%%%%%%%%%%%%%%%%%%%%%%
\begin{figure}
\hspace{0cm}{\includegraphics[scale=0.14]{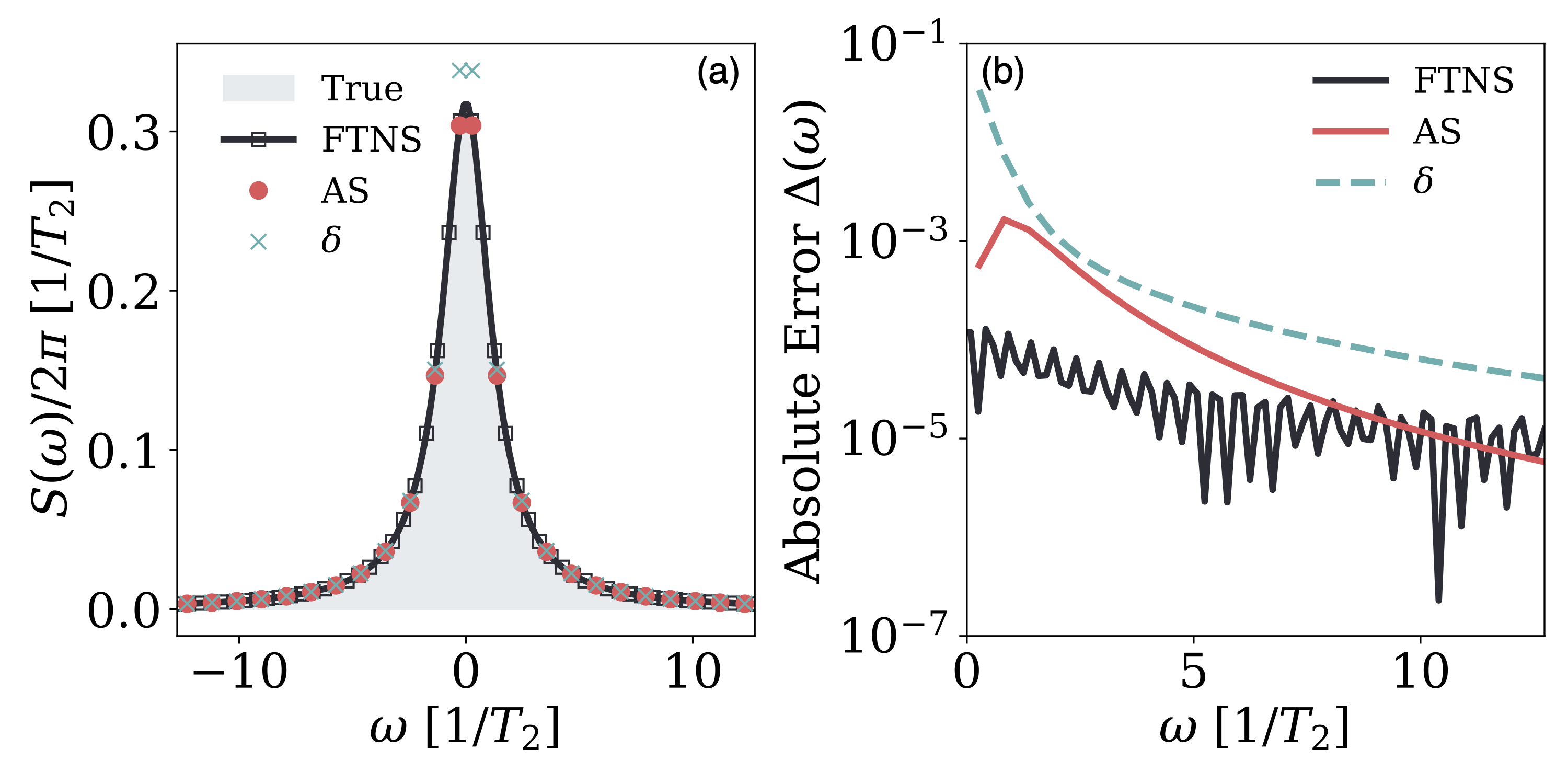}}
\caption{\textbf{Comparison of a simple noise spectrum reconstruction between the FTNS and DDNS methods}: (a) A Lorentzian spectrum and its reconstruction using 32-pulse \'Alvarez-Suter (AS), the 32-pulse single $\delta$-function approximation ($k_\mathrm{c}=1$), and FTNS. \'Alvarez-Suter method: 3662 frequencies have been reconstructed but only selected points have been marked for clarity and only a subset of these fall within the frequency range shown as most fall in the high-frequency regime. FTNS: only frequencies corresponding to the marked \'Alvarez-Suter ones are shown. (b) The absolute error compared to the true spectrum. The spectrum parameters are given in the Methods section~\ref{app-figure-params}. The coherence function contains 1930 points with a resolution of $\delta t/T_2 = 0.00314$ and a final measurement time of $T_\mathrm{max}/T_2 = 6.06$ for the FTNS method, and 3662 points with a minimum measurement time of $T_\mathrm{min}/T_2 = 0.101$ and a final measurement time of $T_\mathrm{max}/T_2 = 368.1$ for the AS and $\delta$-function methods. While DDNS requires many pulses to achieve comparable accuracy and a much longer maximum measurement time for comparable resolution, FTNS outperforms DDNS and uses only free induction decay measurements.}
\label{fig-comparison-i}
\end{figure}
%%%%%%%%%%%%%%%%%%%%%%%%%%%%%%%%%

To illustrate this method for an analytically solvable case, we adopt a Gaussian-shaped noise power spectrum  $S(\omega)=A\mathrm{e}^{-(\omega/\sigma)^2}$. The coherence function of this noise profile can be obtained analytically: 
\begin{equation} \label{eq-w-gaussian}
    C(t)=\exp \left\{-\frac{A}{\sigma}\Bigg[\frac{t\sigma}{2} 
   \text{Erf}\left(\frac{t\sigma}{2}\right
   )+\frac{\mathrm{e}^{-\frac{t^2\sigma^2}{4}}-1}{\sqrt{\pi }}\Bigg]\right\},
\end{equation} 
where $\text{Erf}(z)=2\pi^{-1/2} \int_{0}^{z} \mathrm{e}^{-x^{2}}~d x$ is the Error function. The second derivative of the attenuation function takes the expected Gaussian form, 
\begin{equation}\ddot{\chi}_{\mathrm{FID}}(t) = \frac{A}{\sqrt{2\pi}} e^{-\frac{t^2\sigma^2}{4}},\end{equation}
as does its Fourier transform, 
\begin{equation} \mathcal{F}\Big[\frac{A}{\sqrt{2\pi}} \mathrm{e}^{-\frac{t^2\sigma^2}{4}}\Big]=\frac{A}{\sqrt{2\pi}}\mathrm{e}^{-(\omega/\sigma)^2},\end{equation} 
suggesting $ S_{\text{rec}}(\omega)  = A\mathrm{e}^{-(\omega/\sigma)^2}$. This simple example illustrates that the FTNS method retrieves the original noise spectrum. 

Eq.~\ref{eq-chi-derivative} has important implications for the asymptotic behavior (i.e., $\lim t\rightarrow \infty$) of $\ddot{\chi}_{\mathrm{FID}}(t)$. Specifically, we may invoke the Riemann-Lebesgue lemma~\cite{gradshteyn2014table} for integrable noise spectra --- a physically reasonable assumption. This behavior ensures that $\lim_{t \rightarrow \infty} \ddot{\chi}(t) \rightarrow 0$ and, therefore, that the long-time limit of the attenuation function must grow linearly with time, $\lim_{t \rightarrow \infty} \chi(t) \propto t$. This linear $t$ scaling of $\lim_{t \rightarrow \infty} \chi(t)$ has important implications that we exploit below to provide a theoretically justified and practical approach to inverting experimentally measured coherences, $C(t)$, to well-behaved power spectra, $S(\omega)$.

Translating the above insights into a noise spectroscopy procedure is straightforward. First, one measures the coherence function $C(t)$ from free induction decay by performing Ramsey measurements at various times, yielding an array of coherence values in $[0, T_\text{max}]$ with a sampling interval, or resolution, $\delta t$. One then takes a logarithm of the data and numerically performs a double derivative on the sampled $\chi_{\mathrm{FID}}(t)$ values. A Fourier transform of the resulting data yields the noise spectrum $S(\omega)$. For this, one can employ a discrete Fourier transform or numerical quadrature to obtain equivalent results. 

\subsection*{\textbf{Advantages of FTNS}}

To illustrate the power of the FTNS approach, we assess its ability to reconstruct single-,
\begin{equation} \label{eq-lorentz-1} 
    S(\omega)=\frac{s_0}{1+\left(8 \omega / \omega_\mathrm{c}\right)^{2}},
    \end{equation}
and double-Lorentzian spectra,
\begin{eqnarray}\label{eq-lorentz-2}
    S(\omega) &=& \frac{s_0}{1+\left(8 \omega / \omega_{\mathrm{c},1
}\right)^{2}}+\frac{s_1}{1+8(8[\omega-d] / \omega_{\mathrm{c},2})^{2}} \nonumber \\
&&+\frac{s_1}{1+8(8[\omega+d] / \omega_{\mathrm{c},2})^{2}},
    \end{eqnarray}
that are relevant to bulk~\cite{BarGill2012NatComm} and near-surface~\cite{Romach2015PRL} NV centers, respectively. Here $s_0$ and $s_1$ represent the average coupling strength of the bath to the qubit, and $1/\omega_\mathrm{c}$ is the correlation time of the bath.

Figure 1(a) shows a single Lorentzian peak (grey, shaded) and its spectrum reconstruction using the \'Alvarez-Suter method with a CPMG sequence with 32 $\pi$-pulses total henceforth referred to as the 32-pulse \'Alvarez-Suter (red circles), the single $\delta$-function approximation of the 32-pulse \'Alvarez-Suter (light blue crosses), and FTNS (dark blue line and squares), respectively, in units of $1/T_2$. We assume ideal $\pi$ pulses with perfect fidelity and infinitely short duration throughout the article. In this figure, we aim to show the advantages that each method offers in principle. For this reason, we use a large number of coherence measurements and a long final measurement time. In subsequent figures, we introduce practical considerations to demonstrate how each method can be expected to perform in an experimental setting. The absolute error of the reconstructed spectrum compared to the true spectrum is computed as $\Delta(\omega)=|S(\omega)-S_{\text{rec}}(\omega)|$. As Fig.~1 demonstrates, FTNS outperforms the 32-pulse \'Alvarez-Suter method at low frequencies while only requiring free induction decay measurements. 

We note that for Figs.~1-4, $T_2$ is defined as the inverse of the slope of $\chi(t)$ at sufficiently long times when $\chi(t)$ scales linearly in time. This definition ensures that the $T_2$ time measures the time constant associated with exponential decay, which can only be expected to arise at sufficiently long times. For Figs.~5 and 6, the forms of the spectra make it hard to apply the same definition. In these cases, we employ an alternative definition of $T_2^*$ as the time when the coherence crosses the value $\mathrm{e}^{-1} \approx 0.3678$ for the first time, under a spin echo pulse sequence.

Noting the difficulties of the DDNS approach in accessing the noise spectrum in both $\pi/\tau_\mathrm{max} \leq |\omega^{\rm DDNS}|\leq \pi/\tau_\mathrm{min}$ and $S(\omega \rightarrow 0)$ limits, it is worth considering if and how similar limitations may hinder the FTNS approach. Since our protocol for FTNS relies on the discrete Fourier transform, two quantities determine the highest accessible frequency ($\omega^{\rm FTNS}_{\rm max}$) and its spectral resolution ($\delta \omega^{\rm FTNS}$), which in turn determines the lowest accessible frequency ($\omega^{\rm FTNS}_{\rm min}$): the sampling interval, $\delta t$, of the coherence function measurement, and the total measurement time, $T_\text{max}$. Specifically, $|\omega^{\rm FTNS}_{\rm max}| = \pi/\delta t$ and $|\omega^{\rm FTNS}_{\rm min}| = \delta \omega = 1/T_\text{max}$. While $\delta t$ is determined by limitations of state-of-the-art measuring devices, $T_\text{max}$ depends on the physical problem. Yet, for many cases of physical interest, $\ddot{\chi}(t) \rightarrow 0$ at times earlier than $T_\text{max}$~(Methods section~\ref{app-linear}). This allows one to zero-pad $\ddot{\chi}(t \geq T_\text{max})$ to a new effective $\tilde{T}_{\text{max}} \gg T_\text{max}$, offering sufficient spectral resolution to access $S(\omega \rightarrow 0)$.

Given the importance of $\delta t$ in allowing FTNS to access high frequencies and the analogous role that the minimum delay time, $\tau_\mathrm{min}$, plays in DDNS, we now consider their connection in greater detail. $\tau_\mathrm{min}$ determines the earliest time (after $t=0$) where one can measure the coherence function, i.e., $C(\tau_\mathrm{min})$. Since the discrete Fourier transform requires measurements of $C(t)$ at regular intervals, one might be tempted to assume that $\delta t = \tau_\mathrm{min}$. This need not be the case. After all, for $t \geq \tau_\mathrm{min}$, the measurement interval $\delta t$ is not determined by $\tau_\mathrm{min}$ and can be set such that $\delta t < \tau_\mathrm{min}$. While the resolution of the coherence function measurements at later times can be made as fine-grained as desired, one still needs to perform measurements in $[0, \tau_\mathrm{min}]$ to achieve a consistent $\delta t$ through $[0, T_\text{max}]$. To achieve this, we suggest employing the limit $\omega t \ll 1$, which reveals that the attenuation function behaves as $\chi(t)\approx \alpha t^2+\beta t^4+\gamma t^6$, to fit $C(t)$ at early times~(Methods section~\ref{app-early-time}). This guarantees that FTNS can be implemented even when constrained to the same minimum delay time of dynamical decoupling pulse sequences. 

We are now in a position to illustrate the ability of FTNS to capture a spectrum composed of a sum of Gaussians (Fig.~2(a)), and a double-Lorentzian spectrum (Fig.~2(b)), each compared to a reconstruction using a 32-pulse and a 16-pulse \'Alvarez-Suter subject to the same $\tau_\mathrm{min}$ and $T_\mathrm{max}$ constraint: $C(t\leq T_\mathrm{max}) \geq 0.005$~(Methods section~\ref{app-early-time}). To compare our FTNS results to the best possible \'Alvarez-Suter output, the results shown in Fig.~2 are a combination of results from a 32 (16)-pulse \'Alvarez-Suter procedure, and additional iterations of the 32 (16)-pulse single-$\delta$ function approximation procedure at frequency values between those evaluated by the \'Alvarez-Suter method. We have done this to artificially increase the frequency resolution of the reconstructed \'Alvarez-Suter spectra, although, to our knowledge, this adds a significant experimental expense. Without this, the spectrum reconstruction obtained from the \'Alvarez-Suter procedure would have a much poorer frequency resolution. We also employ this approach to compare against the best possible \'Alvarez-Suter results in Figs.~5 and 6.

%%%%%%%%%%%%%%%%%%%%%%%%%%%%%%%%%
\begin{figure}[t]
\hspace{-.25cm}{\includegraphics[scale=0.1]{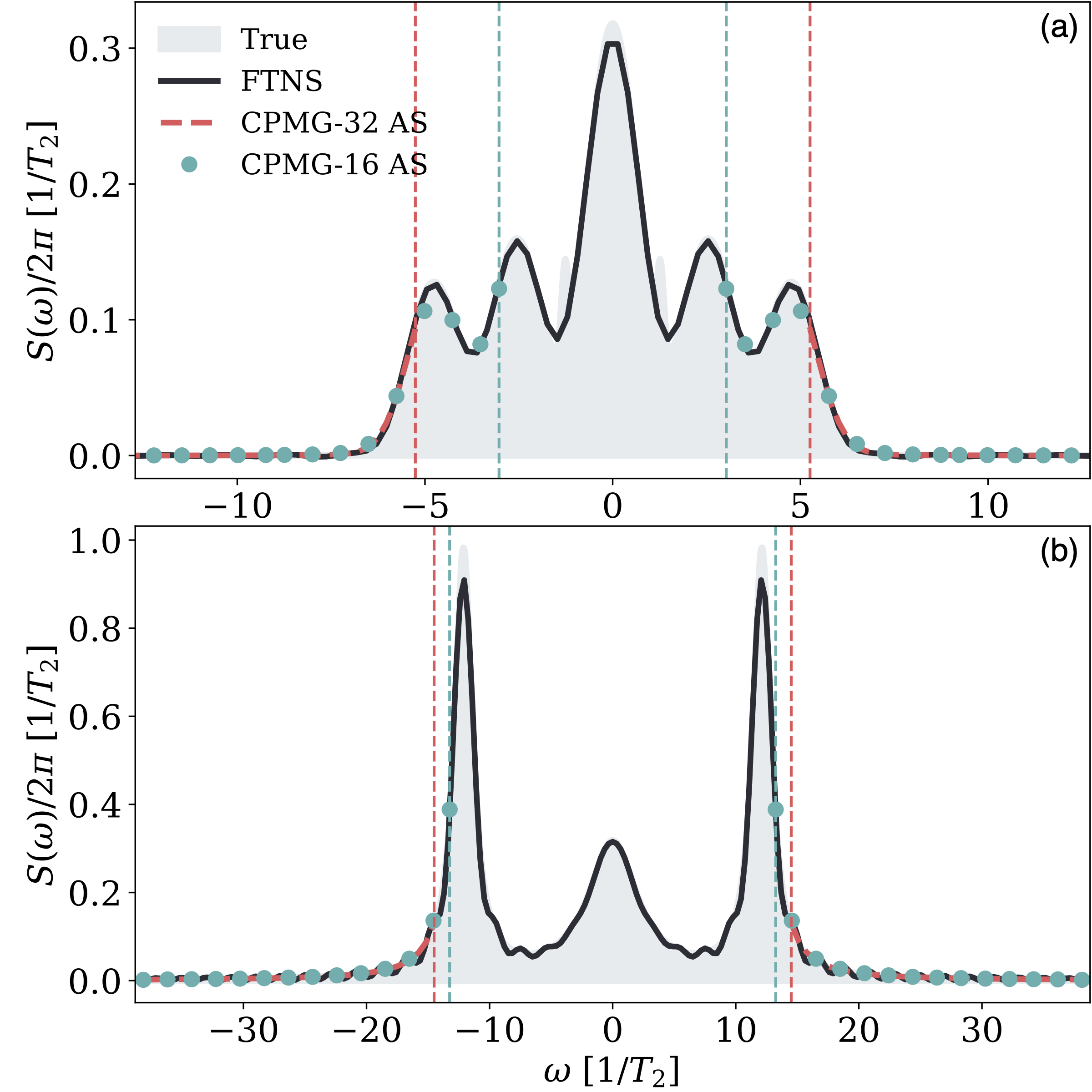}}
\caption{\textbf{Comparison of structured noise spectra reconstructions between the FTNS and DDNS methods}: Examples of two structured noise spectra reconstructed using FTNS (solid dark blue line), a 32-pulse \'Alvarez-Suter (AS) method (red dashed line), and a 16-pulse AS method (light blue dots). The parameters for each spectrum and early-time fitting parameters are given in the Methods section~\ref{app-figure-params}. The coherence function used to implement our FTNS protocol in (a) contains 596 points with a resolution of $\delta t/T_2 = 0.006291$. Only select points have been marked for the 16-pulse AS results, for clarity. For the spectrum in (b) our FTNS coherence function contains 599 points with a resolution of $\delta t/T_2 = 0.005156$. The red and light blue vertical lines indicate the frequency limits up to which the 32-pulse and 16-pulse \'Alvarez-Suter methods (respectively) can reconstruct the spectrum for the given minimum delay time: $|\omega^{\rm DDNS}_\text{min}|\geq \pi/\tau_\mathrm{max}$. Our free induction decay-based FTNS accurately reconstructs the noise spectrum, even in the low-frequency regions containing the main features of the spectra that the AS methods cannot capture.}
\label{fig-comparison-ii}
\end{figure}
%%%%%%%%%%%%%%%%%%%%%%%%%%%%%%%%%

Even in this optimal (but experimentally expensive) \'Alvarez-Suter implementation, both panels of Fig.~2 show that the \'Alvarez-Suter spectra (red dashed line, light blue dots) cannot access their respective $|\omega| < \omega^{\rm DDNS}_{\rm min}$. To go beyond these limits in DDNS, one can employ complex CDD sequences or relax the constraint on the $T_\mathrm{max}$ imposed. In contrast, FTNS only has difficulty resolving the feature at $\omega \sim \pm 1.27\ \mathrm{[}1/T_2\mathrm{]}$ in Fig.~2(a) which is another consequence of the $T_\mathrm{max}$ restriction. Going beyond this maximum measurement time is required to recover this feature using the FTNS method. Further, while reducing the number of pulses used in the \'Alvarez-Suter method allows lower frequencies to be probed, there is a limit to how much the pulse number can be reduced, as at a sufficiently low pulse number, the assumptions underlying the \'Alvarez-Suter method fail to hold. Thus, FTNS uses a simple free induction decay measurement that successfully reconstructs the spectrum in the frequency range that is inaccessible to the DDNS method, giving access to information that would be otherwise lost. 

Since smaller $\delta t$ gives access to higher $\omega^{\rm FTNS}_{\rm max}$ but raises the cost of the experimental procedure, we turn to the trade-off in FTNS accuracy and the sampling interval $\delta t$. Figure 3 depicts FTNS spectrum reconstructions using sets of coherence measurements for a fixed measurement time $T_\text{max}$ with varying $\delta t$. Clearly, increasing the resolution of coherence measurements (i.e., decreasing $\delta t$) improves the accuracy of the FTNS reconstruction, especially at higher frequencies. As expected, even low sampling rates accurately reconstruct the low-frequency part of the spectrum while the high-frequency part can be systematically improved with finer $\delta t$. The ability of FTNS to capture the low-frequency component even at low sampling rates is particularly advantageous for decoherence mitigation purposes as low-frequency noise often dominates decoherence~\cite{Makhlin2003JETPL,Schriefl2006NJP}. Thus, even when measurement resolution is limited, FTNS can be expected to perform well at low frequencies and one can systematically assess its accuracy by monitoring convergence with finer $\delta t$.

%%%%%%%%%%%%%%%%%%%%%%%%%%%%%%%%%
\begin{figure}
\hspace{-.5cm}{\includegraphics[scale=0.14]{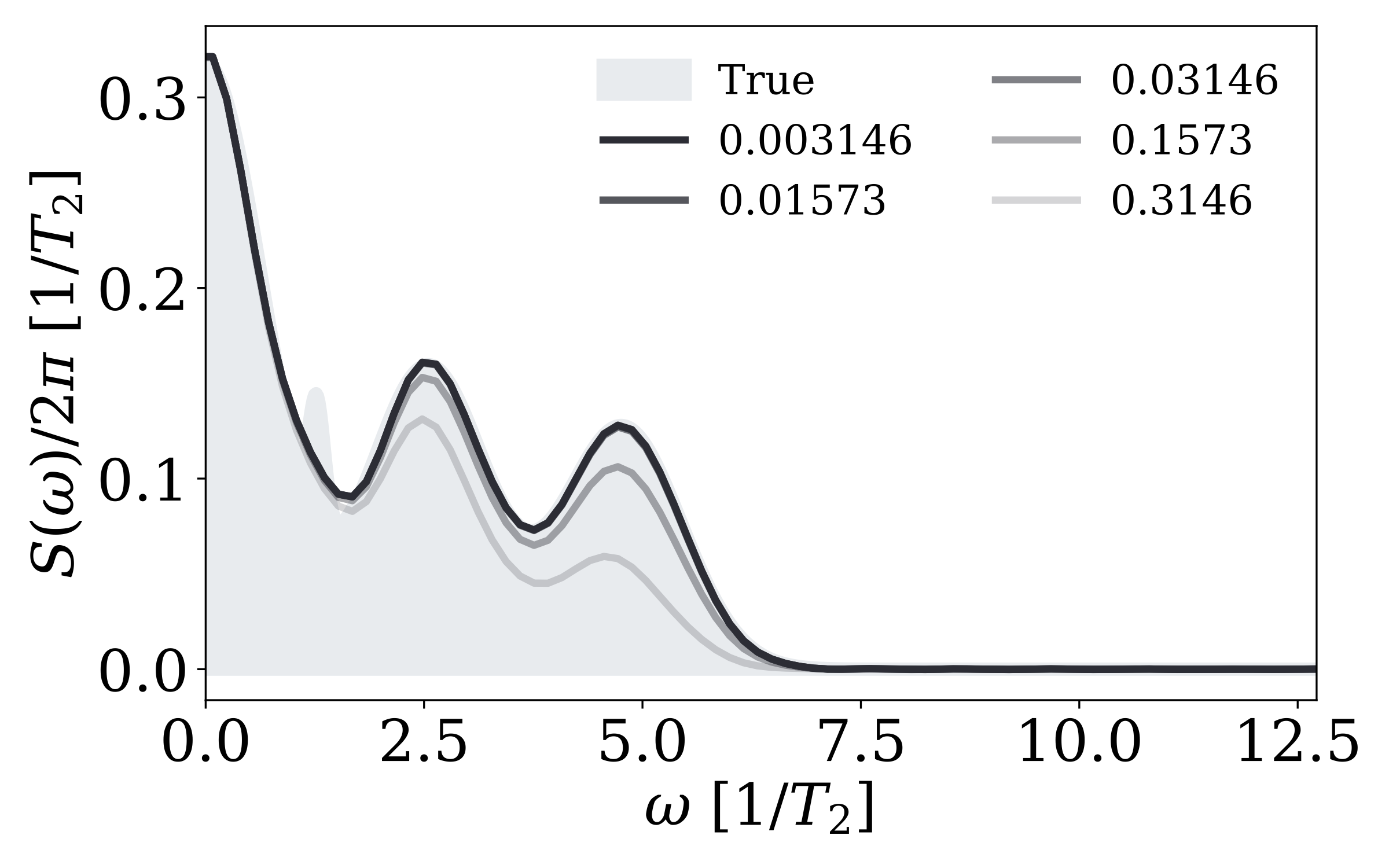}}
\caption{\textbf{Systematic improvement of the noise spectrum reconstructions using FTNS as the sampling interval is decreased}: FTNS reconstructions of the multi-Gaussian spectrum in Fig.~2(a) for different values of the sampling interval, $\delta t$, for a fixed $T_{\text{max}}/T_2=6.291$ (other parameters are the same as Fig.~2(a)). The legend shows the values of $\delta t/T_2$ used to obtain the FTNS results shown.} 
\label{fig-resolution}
\end{figure}
%%%%%%%%%%%%%%%%%%%%%%%%%%%%%%%%%

\subsection*{\textbf{Robustness against errors}} \label{sec-spam}

Since FTNS requires performing two numerical derivatives, it is sensitive to errors that occur during the initialization and measurement phase (e.g., state preparation and measurement (SPAM) errors, and statistical uncertainty due to finite measurement repetition) of the protocol. There are multiple sources of errors that can compromise the measured value of the coherence function at a given time. These include background and shot noises, and imperfect fidelity of the applied pulses~\cite{Tyryshkin2010,Souza2011PRL}. In optical setups, photon losses can also reduce the number of effective measurements. Nevertheless, various methods to perform controlled numerical derivatives of noisy data are available~\cite{Chartrand2011IAM,VanBreugel2020ieee}. As an example, here we utilize a simple denoising method that mitigates the effect of noise and preserves all the advantages of FTNS even on structured noise spectra. Figure 4 shows examples of FTNS spectra reconstructed from artificially noisy data corresponding to an effective fixed measurement error of $0.1\%$ of the range (difference between maximum and minimum) of the coherence function (for examples of the same reconstruction with higher error rates, see Fig.~7). 
This Gaussian-distributed statistical noise is meant to mimic all uncorrelated errors in coherence measurements, including readout errors, which can be grouped with the shot noise. By increasing the number, $N$, of repetitions (i.e., individual measurements of the spin coherence function via ensemble readout at each control condition), the statistical variation of the signal diminishes as $1/\sqrt{N}$. We perform linear fitting of $\chi(t)$ at long times (consistent with the asymptotic behavior of $\ddot{\chi}(t)$ established using the Riemann-Lebesgue lemma in the section on the Theoretical Description) and apply low-pass filters to recover the approximate noise spectrum, which shows good agreement with the true spectrum, revealing its essential features~(see Methods section~\ref{app-linear} for details). Hence, our simple free induction decay-based noise spectroscopy approach can semi-quantitatively recover the frequency and the height of the peaks of the noise spectrum, which constitute the minimum required spectral information to design effective filter functions to mitigate decoherence from a dynamical decoupling perspective. 

%%%%%%%%%%%%%%%%%%%%%%%%%%%%%%%%%
\begin{figure}[b]
\hspace{0cm}{\includegraphics[scale=0.1]{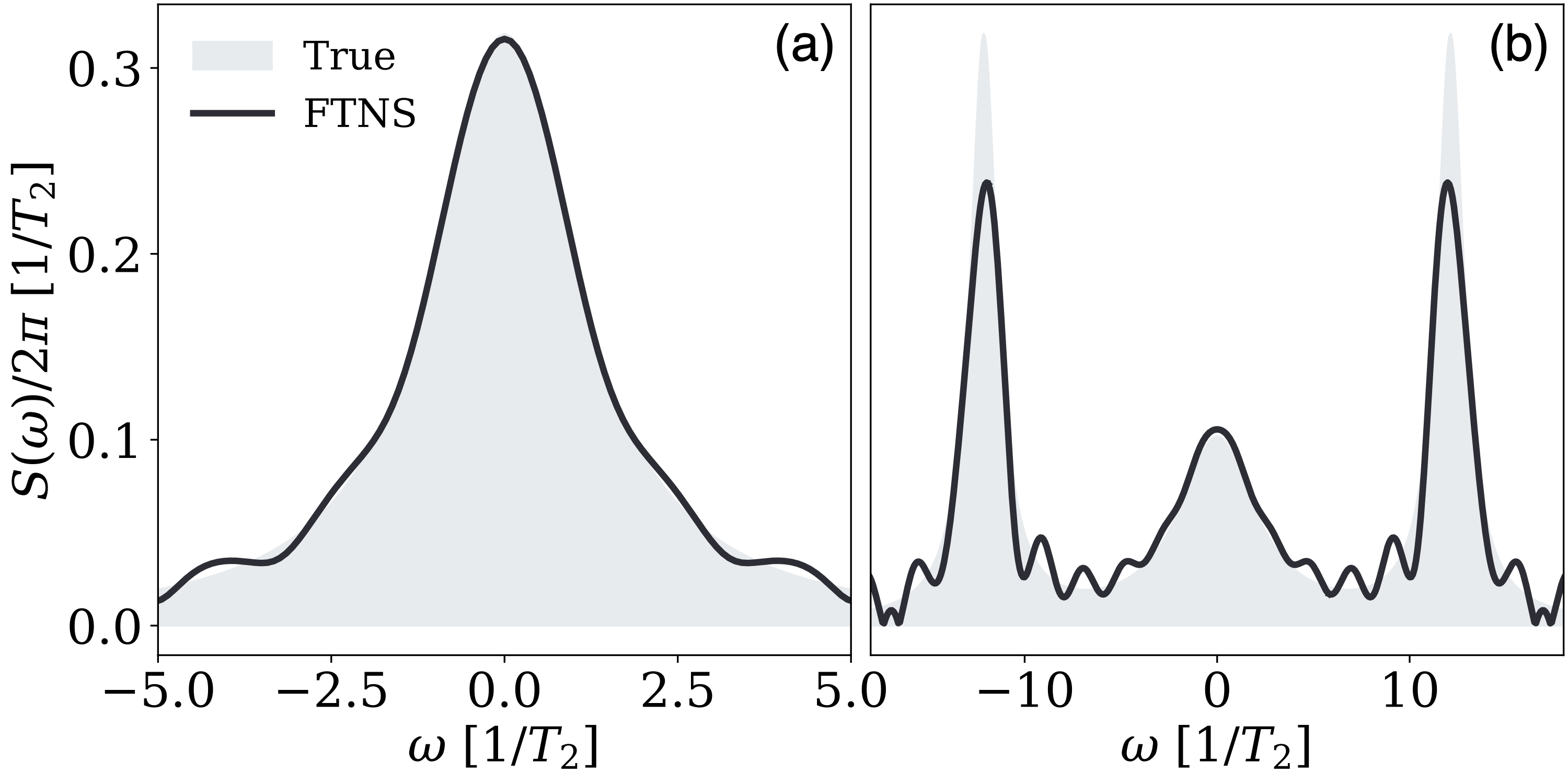}}
\caption{\textbf{Noise spectrum reconstructions using the FTNS method under simulated measurement noise}: Spectrum reconstruction using FTNS assuming $0.1\%$ measurement error in the coherence measurements for the same spectra shown in Figs.~1(a) and 2(b) in panels (a) and (b), respectively (with the same parameters). Even subject to SPAM errors, utilizing simple denoising techniques allows FTNS to quantitatively capture the height and frequency of the peaks near $\omega = 0$ and qualitatively obtain the peaks at higher frequencies, in the spectrum.}
\label{fig-noisy}
\end{figure}
%%%%%%%%%%%%%%%%%%%%%%%%%%%%%%%%%

While our analysis thus far accounts for the theoretical constraints of the discrete Fourier transform, the experimental feasibility of FTNS is sensitive to $\delta t$ and the minimum delay time $\tau_\mathrm{min}$. Controlling these parameters requires flexibility in pulse design, which varies depending on the platform. For instance, solid-state spins can be controlled either optically with pulses that range from a few ps up to 1 ns~\cite{Press2010NatPhot,Vezvaee2023PRX,Bodey2019npj,Takou2021PRB} or via microwave pulses as short as 12 ns \cite{Nguyen2019PRB}. Such pulses allow for ns-scale minimum delay times, $\tau_\mathrm{min}$, between pulses. Furthermore, tuning the sampling interval $\delta t$ to ps-order precision is also achievable. For example, from an implementation perspective, two microwave pulses with a controllable delay can be generated with an arbitrary waveform generator and the delay between two optical pulses can be easily controlled by varying the length of an arm in the Michelson interferometer. To illustrate how these timescales satisfy our FTNS requirements, consider the $\delta t$ required to reconstruct the double-Lorentzian spectrum shown in Fig.~2(b) in the NV center parameter regime~\cite{BarGill2012NatComm,Bauch2018PRX,Bauch2020PRB,Taylor2008Nature} with $T_2=1.32~\mu $s. To obtain the accurate reconstruction shown in Fig.~2(b), one would need to measure 599 points with $\delta t\approx 7$~ns. Thus, sufficiently high-resolution measurements that faithfully reconstruct various noise spectra can be comfortably performed with experimentally available technology.

Since the feasibility of FTNS also relies on the ability to sufficiently reduce statistical noise ($\sim$0.1\%) within a reasonable time, we now consider what current technology can afford. Each point on the coherence curve arises from a Ramsey measurement at a given time, repeated multiple times to construct the single-qubit ensemble average. To minimize statistical noise associated with finite sampling, the repetition rate of such experiments needs to be sufficiently high. To estimate the time required for the FTNS experiment, we multiply the time it takes to perform a single-shot experiment on a single data point along the coherence curve by the number of independent measurements required to bring the statistical noise level to the desired threshold, under the assumption that independent random Gaussian fluctuations can satisfactorily model the combined effects of the sources of statistical noise. We apply this protocol to find the time required on available experimental setups to bring the statistical error to the 0.1\% value assumed in Fig.~4 in the measured coherence of NV centers with $T_2 \sim$ few $\mu s$~\cite{BarGill2012NatComm,Bauch2018PRX,Bauch2020PRB}, with access to nanosecond microwave pulses. This 0.1\% error requires that each point along the coherence curve be measured $\sim 10^6$ times. A single measurement takes $\sim 10~\mu$s (including the initialization and readout). Thus, requiring $\sim 100$ data points along $C(t)$ takes $\sim 2.8$ hours (assuming a modest photon collection rate of $\sim 10\%$). Importantly, these numbers represent a conservative estimate and can be expected to improve significantly with better photon collection rates or through non-uniform sampling techniques~\cite{Fessler2003,Wernecke1977,Kazimierczuk2011,Holland2011,Jaravine2006}. Further, by requiring only two $\pi/2$ pulses, free induction decay-based FTNS avoids measurement error arising from imperfect pulses, which accrues significantly in large dynamical decoupling pulse sequences with many $\pi$ pulses.

\subsection*{\textbf{Spin-Echo FTNS}} \label{sec-se-ftns}

The FTNS protocol that we have discussed so far employs free induction decay coherence measurements that decay sufficiently slowly so as to allow enough measurements of the coherence curve to support a well-behaved Fourier transform. A fast decaying behavior can arise from a sharply peaked low-frequency noise component at $\omega=0$. In such scenarios, it is customary to utilize a single $\pi$-pulse spin echo (SE) sequence to remove the effect of the low-frequency component of $S(\omega)$ to prolong the $T_2$ time. It would therefore be beneficial to provide a one-to-one map and a noise spectroscopy protocol to perform FTNS based on the spin echo sequence. Below we derive this one-to-one map and offer a practical protocol for spin echo-based FTNS. We further show that while spin echo-based FTNS tends to be less accurate at low frequencies, it can outperform the free induction decay-based method at higher frequencies, especially in the presence of strong low-frequency noise. In addition, the spin echo-based method enables the reconstruction of $1/f$-type spectra which is not possible using the free induction decay-based method.

The filter function of the spin echo sequence is $F_\text{SE}(\omega,t)=(16/\omega^2)\sin^4(\omega t/4)$. Following similar steps to those used for free induction decay, we take a double derivative of the spin echo attenuation function to find, 
\begin{equation}
 \ddot{\chi}_{\mathrm{SE}}(t) =\frac{1}{2\pi} \int_{-\infty}^{\infty} d \omega S(\omega)[\cos (\omega t/2)-\cos ( \omega t)], 
\end{equation}
and therefore, 
\begin{equation} \label{eq-SE-FTNS}
  \sqrt{2\pi}  \mathcal{F}[\ddot{\chi}_{\mathrm{SE}}(t)]= 2M(2\omega),
\end{equation}
where $M(\omega)\equiv S(\omega)-S(\omega/2)/2$. $M(\omega)$ corresponds to an array of measurements that can be performed at regular values of $\omega_n \in [0, \delta \omega, 2\delta \omega, ..., n_{\rm max}\delta \omega]$, where $n \in \mathbb{N}$, separated by an interval $\delta \omega$ that are accessible via the Fourier transform of the second derivative of the spin echo coherence function. Hence, we write $M(\omega_n) = M(n \times \delta \omega)$. 

While the map in Eq.~\eqref{eq-SE-FTNS} provides the spectral function $S(\omega)$, it also contains an unwanted part, $S(2\omega)$, which we need to isolate and remove. We introduce a simple recursive method that allows us to extract the spectral function $S(\omega)$ from our spin echo-based FTNS results:
\begin{eqnarray}
    S(2n\delta \omega)&=&M(2n\delta \omega)+\frac{1}{2}S(n\delta \omega), \label{eq-even}\\
    S((2n+1)\delta \omega)&=&M((2n+1)\delta \omega) \nonumber \\
    &&+\frac{1}{4}(S((n+1)\delta \omega)+S(n\delta \omega)). \label{eq-odd}
\end{eqnarray}

To arrive at this result, we first exploit the fact that $S(0)=2 M(0)$. While the Fourier transform does not give access to $M(\omega = 0)$, it can be interpolated. We then approximate $S(n \delta \omega/2)$ for odd $n$ as the arithmetic average of two adjacent points, $S(n \delta \omega/2)\approx(1/2)(S((n-1) \delta \omega/2)+S((n+1) \delta \omega/2))$, allowing us to find $S(\delta \omega)=(4/3)(M(\delta \omega)-S(0)/4)$.

%%%%%%%%%%%%%%%%%%%%%%%%%%%%%%%%
\begin{figure}
\hspace{-.2cm}{\includegraphics[scale=0.14]{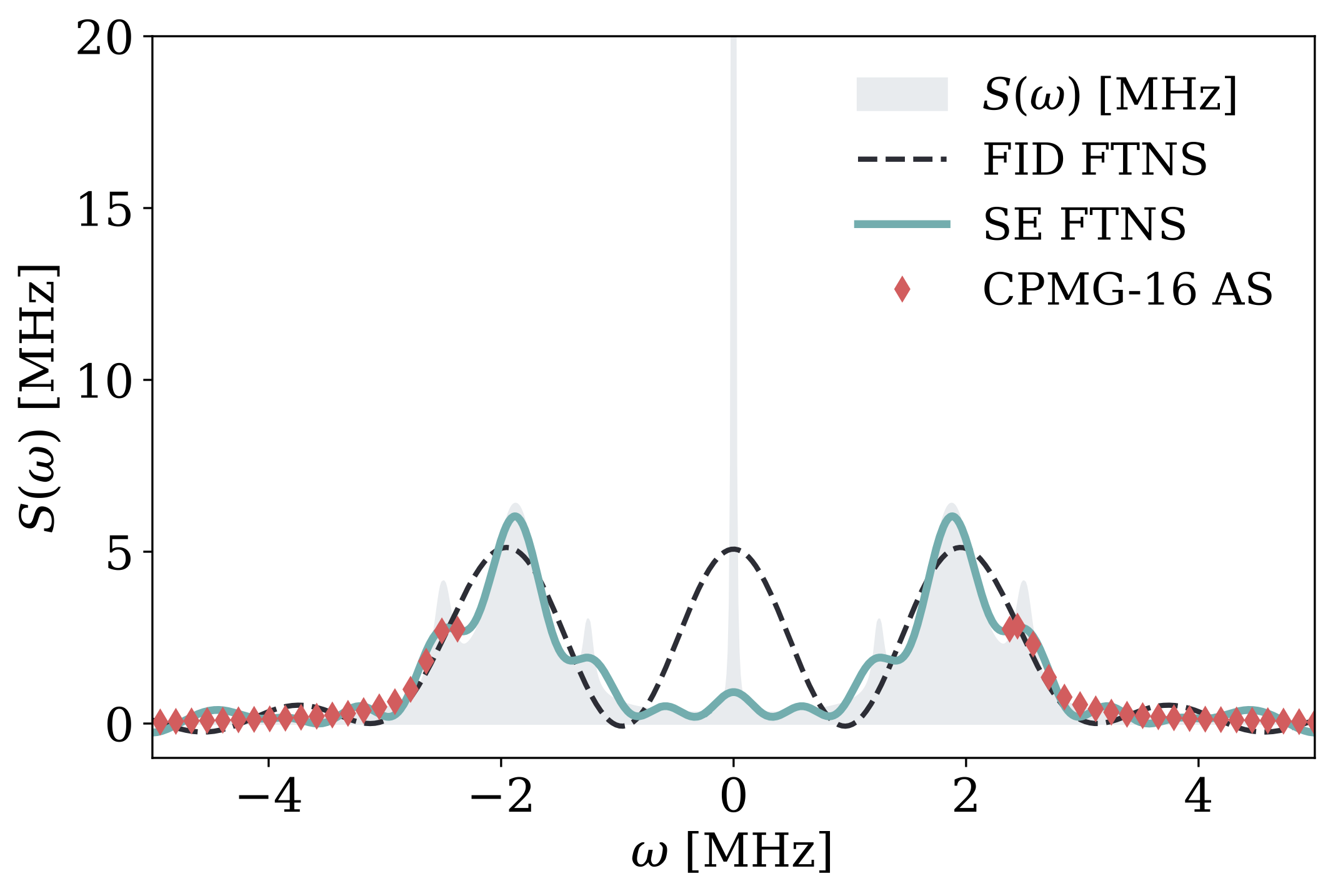}}
\caption{\textbf{Comparison of structured noise spectrum reconstruction between the SE FTNS, FID FTNS, and DDNS methods}: Performance of spin echo (SE) FTNS, free induction decay (FID) FTNS, and 16-pulse AS for a sharply peaked low-frequency double-Lorentzian spectrum, with $T_2^* = 1.646$ (as defined in the main text). Only every other point in the AS reconstruction has been plotted for clarity. The FTNS methods identify some of the low-frequency and high-frequency components of the spectrum. In particular, while the FID FTNS identifies the presence of peaks at $\omega \sim 0$ and at $\omega \sim 2$, it cannot capture their details accurately. On the other hand, SE FTNS is blind to the sharp peak at zero frequency but resolves the finer structure in the higher frequency peaks in the spectrum. In contrast, the DDNS method again fails to reconstruct the relevant low-frequency regions of the spectrum due to the restrictions imposed by $\tau_\mathrm{max}$.}
\label{fig-SE}
\end{figure}
%%%%%%%%%%%%%%%%%%%%%%%%%%%%%%%%

To see the performance of the spin echo version of the FTNS, we apply this protocol to an experimentally inspired, highly peaked low-frequency double-Lorentzian spectrum given by 
\begin{eqnarray}\label{eq-lorentz-SE}
    && S(\omega) = \frac{s_0}{1+\left(\frac{8 \omega }{ \omega_{\mathrm{c},0}}\right)^{2}} \nonumber \\
    && \quad + \sum_{i=1}^3 \frac{s_i}{1+8\Big(\frac{8[\omega-d_i]}{ \omega_{\mathrm{c},i}} \Big)^{2}}  +\frac{s_i}{1+8\Big(\frac{8[\omega+d_i] }{ \omega_{\mathrm{c},i}}\Big)^{2}},
\end{eqnarray}
with parameter values as given in Methods section~\ref{app-figure-params}. For this spectrum, the maximum measurement time (obtained by imposing the condition that $C(t)>0.005$ at all times) under the spin echo sequence is about 5.5 times longer than that under free induction decay, corresponding to a moderate scenario in which we envision the spin echo-based FTNS offering an advantage. Figure 5 shows that the spin echo-based FTNS entirely misses the presence of the narrow low-frequency peak but faithfully captures the tri-peak structure of the mid-frequency feature in the $1-5$ MHz range. Instead, the free induction decay-based FTNS identifies a peak at low frequency but is unable to capture any structure for the mid-frequency peak. The inability of the spin echo method to capture the sharp feature at $\omega \approx 0$ is likely because the application of the single $\pi$-pulse removes the inhomogeneous ($\omega \rightarrow 0$) contribution in the power spectrum. In contrast, the inability of the free induction decay method to capture the mid-frequency features likely arises from the fast decay of the signal and the stringent limit on the measurement time. What is most remarkable is that in such systems with a dominant inhomogeneous contribution, the spin echo-based method can resolve the finer structure in the higher frequency peaks compared to the free induction decay method. This illustrates a distinct benefit arising from an increased coherence time on the performance of the FTNS method. In contrast to FTNS, 16-pulse DDNS is again unable to capture many of the prominent features of the spectrum. What is more, one needs to resort to a 16-pulse sequence instead of a 32-pulse sequence to at least partially reconstruct the prominent feature of the spectrum for both Figs.~5 and 6. Thus, both free induction decay and spin echo FTNS methods perform better than DDNS, capturing an informative description of both the central peak and the higher frequency components.

Another important advantage of spin echo FTNS is that it enables one to reconstruct spectra that scale as $1/\omega^n$ at low frequencies (termed $1/f$ spectra), which are observed in many relevant systems~\cite{Wise2021PRX,paladino20141}. The one-to-one correspondence between the $S(\omega)$ and $\chi_\mathrm{SE}(t)$ through the spin echo FTNS in Eq.~\eqref{eq-SE-FTNS} provides a unique pathway to analytically derive an expression for the attenuation function of a $1/f$-type spectrum. Namely, for $S_\mathrm{1/f}(\omega)=\mathcal{A}/\omega^n$, we find
\begin{equation}
    \chi_{\rm{SE},1/f}(t) =\mathcal{Y}_nt^{n+1}, \label{eq-chi_1overf_main}
\end{equation}
where $n$ is a positive value less than 3, $\Gamma(\cdot)$ is the gamma function (see Methods section~\ref{app-SE-1overf}), and the coefficient function $\mathcal{Y}_n$ is given by
\begin{equation} \label{eq-y_n}
\begin{cases}
    \mathcal{A}\log(2)/4\pi  & n=1, \\
   \mathcal{A}/24, & n=2, \\
   -\mathcal{A}\pi^{-1}\left(1-2^{1-n}\right) \sin \left(\frac{\pi n}{2}\right) \Gamma (-n-1), & \rm{otherwise}.
\end{cases}
\end{equation}
The focus on $n<3$ is motivated by experiments in spin qubits~\cite{Medford2012PRL}. These solutions reveal that $\chi_\mathrm{SE}(t)$ due to a $1/f$ spectrum is proportional to $t^{n+1}$. In fact, the same asymptotic time dependence for the $1/f$ spectrum has been approximately obtained for various pulse sequences, including CPMG~\cite{Cywinski2008PRB} and used to analyze the output of noisy dynamical decoupling data~\cite{Medford2012PRL}. Thus, $1/f$ noise spectra stand in contrast to integrable counterparts that remain finite over the entire frequency range (e.g., Gaussian and Lorentzian peaks), for which we used the Riemann-Lebesgue lemma to demonstrate that $\lim_{t \rightarrow \infty} \ddot{\chi}(t) \rightarrow 0$, implying that $\chi(t)$ is proportional to $t$ at long times (see Methods section~\ref{app-linear}). These distinctly different behaviors of finite versus $1/f$-type spectra enable one to distinguish the two, even at the level of the asymptotic scaling of the response function, $\chi(t)$. Specifically, by performing a polynomial fit of the $t$ dependence of the measured $\chi(t)$, one can obtain the parameters that characterize the $1/f$ response needed to fully reconstruct $S(\omega)$ (see Methods section~\ref{app-SE-1overf}). Finally, we note that for $n \geq 3$, the integral expression for $\chi(t)$ under spin echo, as given by Eq.~\ref{eq-coherence-def}, diverges. This suggests that pulse sequences with higher numbers of pulses (and therefore higher orders of $\sin(\omega t)$ in the filter function for CPMG, for example) need to be applied to probe such noise spectra. In principle, the procedure to arrive at the spin echo FTNS can be repeated for such alternate pulse sequences for the ability to probe $1/f$ noise spectra beyond $n=3$. Furthermore, previous work has shown that multi-pulse CPMG sequences with an even number of pulses have a filter function that scales as $F(\omega t) \propto \omega^4$ at low-frequencies~\cite{Cywinski2008PRB, Medford2012PRL}, implying that $\chi(t)$ is finite for $n<5$. When the number of pulses is odd, the filter function scales as $F(\omega t) \propto \omega^2$ at low-frequencies~\cite{Cywinski2008PRB, Medford2012PRL}, implying that $\chi(t)$ is finite for $n<3$.

%%%%%%%%%%%%%%%%%%%%%%%%%%%%%%%%
\begin{figure}[t]
\hspace{-.2cm}{\includegraphics[scale=0.14]{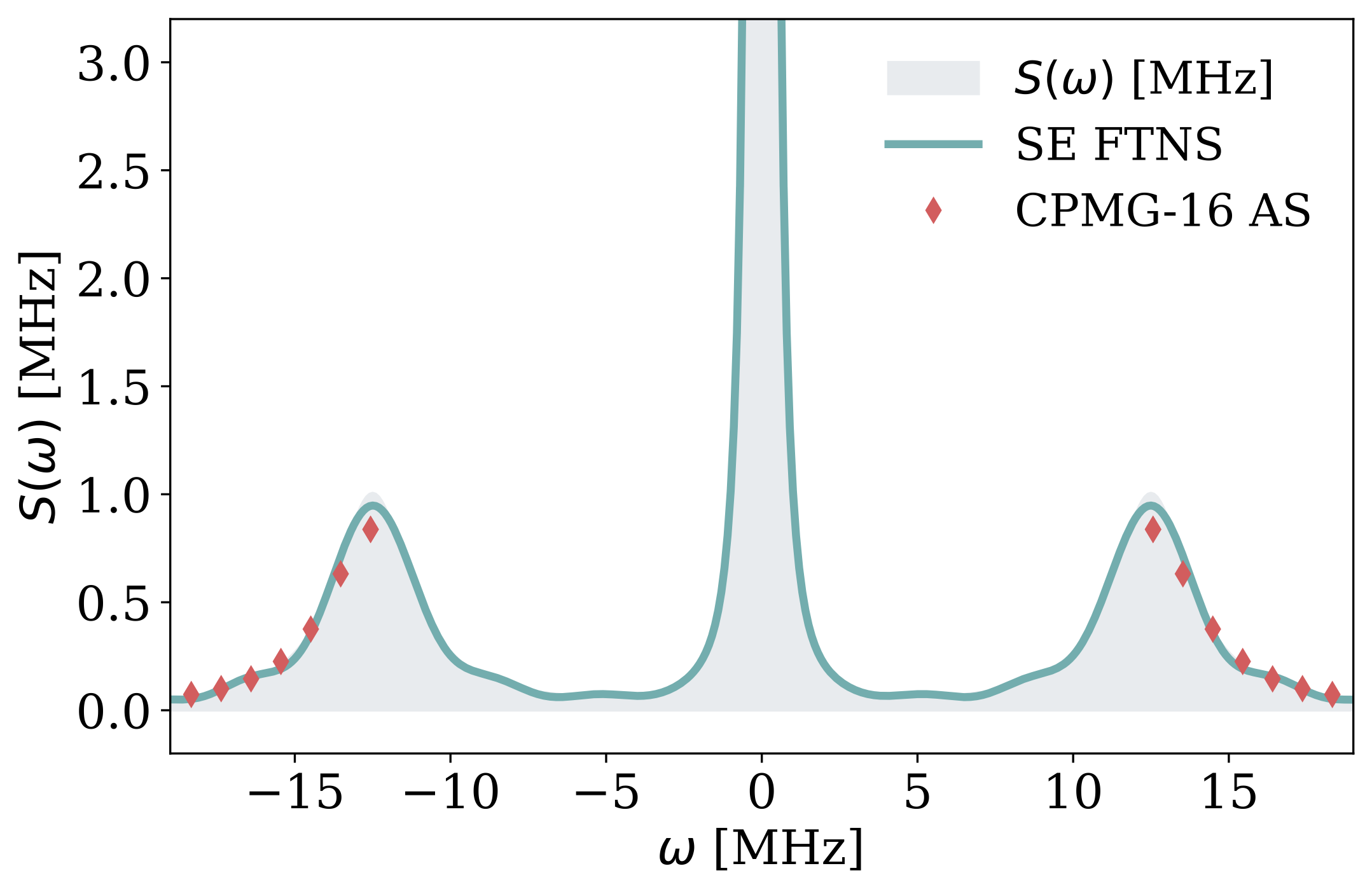}}
\caption{\textbf{Comparison of $1/f$-type noise spectrum reconstruction between the SE FTNS and DDNS methods}: Performance of the spin echo FTNS and AS methods in reconstructing $1/f$-type noise (as given in Eq.~\ref{eq:1overf_Sw_methods}), with $T_2^* = 2.524$. The same coherence measurement constraints of $t_\mathrm{min}=0.01$ and $t_\mathrm{max}=4$ are imposed on both methods, leading to a 300-point reconstruction for the spin echo FTNS method and a 330-point reconstruction for the AS method. The AS method is again unable to reconstruct relevant low-frequency characteristics of the spectrum due to the limitations in maximum coherence measurement time, and the majority of the reconstructed points lie in the higher frequency regions, for which zeros are evaluated. In contrast, the spin echo FTNS method accurately reconstructs both the low frequency $1/f$ behavior and the positions and widths of the finite peaks located at high frequencies, with high resolution.}
\label{fig-1_f}
\end{figure}
%%%%%%%%%%%%%%%%%%%%%%%%%%%%%%%%

While this analysis shows that it is possible to distinguish the presence and $\omega$-scaling of diverging contributions to the power spectra, we now demonstrate that our spin echo-based FTNS also enables us to quantitatively reconstruct both diverging and well-behaved contributions to the power spectrum. Specifically, in Fig.~6, we interrogate the ability of this procedure to disentangle and reconstruct a complex noise spectrum consisting of $1/f$-type and always-finite (Lorentzian) contributions. For the spin echo FTNS, we process the coherence data as outlined in Methods section~\ref{app-SE-1overf}. The $1/f$ parameters that we extract, even under the experimentally motivated constraint of setting $C(t_\mathrm{max}) > 0.005$, agree well with the true spectrum parameters while the 16-pulse DDNS method is again able to retrieve only part of the Lorentzian contributions to the spectrum. What is more, the total reconstructed spectrum obtained \textit{simply through a spin echo measurement} faithfully captures both the positions and widths of the high-frequency peaks and the low-frequency $1/\omega$ behavior. Importantly, as a consequence of the Riemann-Lebesgue lemma, this protocol for $1/f$ noise spectroscopy can be used to disentangle spectra consisting of combinations of $1/f$ form and any other form $f(\omega)$, as long as $\int_{-\infty}^\infty \vert f(\omega)\vert d\omega < \infty$. Hence, the combination of free induction decay- and spin echo-based FTNS allows one to address complex spectra consisting of combinations of the most commonly encountered functional forms in physical systems.
%, while offering advantages over other noise spectroscopy protocols, as emphasized earlier. This sets our method apart as being both generally applicable and accurate over a wide frequency range, while being simple to implement.

%%%%%%%%%%%%%%%%%%%%%%%%%%%%%%%%%
%%%%%%%%%%%%%%%%%%%%%%%%%%%%%%%%%
%%%%%%%%%%%%%%%%%%%%%%%%%%%%%%%%%

\subsection*{\textbf{Summary and outlook}} \label{sec-conclusion}

We conclude the comparison of both versions of FTNS and state-of-the-art DDNS with a few general remarks regarding their applicability. First, the information about low-frequency components in $S(\omega)$ is encoded in the long-time behavior of the coherence function $C(t)$. However, since one cannot measure arbitrarily small values of the coherence function $C(t)$ at long times, we set a measurement cut-off of $C(t)>0.005$ for all methods considered, i.e., \'Alvarez-Suter DDNS and free induction decay- and spin echo-based FTNS (see Methods section~\ref{app-spam}). This limits the number of points that can be reconstructed via DDNS and can also lead to the numerical instability of the Fourier transform. While this can lead to poor resolution of the reconstructed spectrum for the DDNS method, we have shown that FTNS can recover the prominent features of the noise spectrum, albeit at the cost of sometimes introducing unphysical oscillations that can be tamed with more extensive measurements. Second, one can invert higher-order dynamical decoupling sequences via the FTNS method and extract the noise power spectrum from the resulting Fourier transform via a similar iterative approach as that outlined for our spin echo-based FTNS. We show that in systems where the coherence time is greatly improved through the application of a spin echo sequence, our spin echo-based FTNS can outperform the free induction decay-based method in reconstructing high-frequency spectral features, at the cost of discarding information about the zero-frequency spectral features. Thus, while the accuracy of DDNS  requires a large number of pulses, simple free induction decay and spin echo measurements suffice for our FTNS procedures. As we have shown, our spin echo FTNS even enables one to disentangle and accurately reconstruct spectra containing mixtures of $1/f$-type and always-finite contributions. In cases where the spectrum consists of only always-finite contributions, simple free induction decay measurements provide the same spectral information. Hence, even for systems whose free induction decay coherence decays rapidly, FTNS offers significant advantages over DDNS in terms of resolution and simplicity of implementation.

In summary, we have introduced a noise spectroscopy method that significantly outperforms current DDNS methods and is significantly easier to implement from both experimental and theoretical perspectives. Our work demonstrates the existence of a direct one-to-one invertible map between the pure dephasing coherence function within the filter function formalism and the noise power spectrum. Noting that current technology allows one to minimize measurement and statistical errors, it is clear that FTNS provides a promising route to accurately and inexpensively measure noise power spectra. Our FTNS performs efficiently when free induction decay occurs sufficiently slowly, as in trapped-ion systems~\cite{Bruzewicz2019APR,Ruster2016APB}. We have further developed a spin echo FTNS protocol that enables the characterization of fast decaying systems exhibiting $1/f$-type noise, allowing us to reconstruct even spectra dominated by strong inhomogeneous contributions, as in most NV centers~\cite{BarGill2012NatComm,Romach2015PRL}. Therefore, our FTNS protocol should be applicable to a wide range of quantum platforms and can be utilized as a powerful tool to deduce information about the environmental interactions that lead to the decoherence of qubits or quantum sensors.

\section*{Methods}

\subsection{\textbf{Figure parameters}} \label{app-figure-params}

Here, we list the spectrum parameters for each figure. 

For Fig.~1, the Lorentzian spectrum is given by Eq.~\eqref{eq-lorentz-1} with $s_0=2.000/T_2$ and $\omega_\mathrm{c}=10.186/T_2$.

For Fig.~2(a), we employed a combination of four Gaussians,
\begin{equation}
    S(\omega)=\sum_i A_i e^{-(\omega -\mu_i)^2/\sigma_i^2},
\end{equation} 
where $A_i \in \{1.998,0.3995,0.7990,0.9988\}/T_2$, $\sigma_i \in \{0.9537,0.1272,0.9537,0.9537\}/T_2$, and $\mu_i\in \{0.000,1.272,4.769,2.543\}/T_2$. The parameters we obtained from the early-time fitting are $\{\kappa^{(0)},\kappa^{(1)},\kappa^{(2)}\}=\{0.7667,-0.5827,0.3903\}$.
For Fig.~2(b), we used a double Lorentzian given by Eq.~\eqref{eq-lorentz-2} with parameters $s_0=1.939/T_2$, $\omega_{\mathrm{c},1} = \omega_{\mathrm{c},2}=19.39/T_2$, $d=12.12/T_2$, and $s_1=6.093/T_2$. We obtained the following early-time parameters $\{\kappa^{(0)},\kappa^{(1)},\kappa^{(2)}\}=\{3.777,-117.1,59380\}$.

For Fig.~5, the spectrum is given by Eq.~\ref{eq-lorentz-SE}, with parameter values $s_i \in \{150\pi,2\pi,\pi,2\}$, $\omega_{\mathrm{c},i} \in \{0.02,6,2,1\}$, and $d_i \in \{15/8,20/8,10/8\}$.

For Fig.~6, $S(\omega)$ is given by 
\begin{equation}
    S(\omega) = \frac{\mathcal{A}}{\omega^n} + \frac{B}{1+\left(\frac{\omega-d}{\omega_\mathrm{c}}\right)^2}+\frac{B}{1+\left(\frac{\omega+d}{\omega_\mathrm{c}}\right)^2},
\end{equation}
with parameter values $\mathcal{A}=1, \,\,B=1, \,\,n=2.5, \,\,\omega_\mathrm{c} = 1.5, \,\,d=12.5$.

\subsection{\textbf{Early time measurement fitting}} \label{app-early-time}

Here we report the parameters we have used for the figures in the main text. However, before turning to each figure, we first detail the fitting procedure we employ to access the short-time values of the coherence function, $C(t)$, when the measurement resolution is smaller than the minimum delay time of the $\pi/2$ pulses, i.e., $\delta t < \tau$. 

As discussed in the main text, since our sampling interval $\delta t$ is smaller than the minimum delay time $\tau$, we obtain effective coherence function measurements at early times $[0,\tau]$ by employing the small $\omega t$ limit of the free induction decay attenuation function, $\chi_{\mathrm{FID}}(t)$. For early times (i.e., when $\omega t \ll 1$), one can expand $\chi_{\mathrm{FID}}(t)$ as
\begin{equation} \label{eq:chi-short-wt-expansion-FID}
\begin{split}
    \chi_{\mathrm{FID}}(t) &= \frac{1}{\pi} \int_{-\infty}^{\infty} d \omega~ \frac{1}{\omega^2}S(\omega)\sin^2\Big( \frac{\omega t}{2}\Big)\\
    &\approx \frac{1}{\pi} \int_{-\infty}^{\infty} d \omega~ \frac{1}{\omega^2}S(\omega)\Big[ \frac{(\omega t)^2}{2^2}-\frac{(\omega t)^4}{2^4 \cdot 3}+\frac{2(\omega t)^6}{2^6 \cdot 45}\Big]\\
    &\equiv \kappa^{(0)} t^2+\kappa^{(1)} t^4+\kappa^{(2)} t^6,
\end{split}
\end{equation}
where
\begin{eqnarray}
    \kappa^{(0)} &=& \frac{1}{2^2\pi} \int_{-\infty}^{\infty} d \omega~ S(\omega),\\
    \kappa^{(1)} &=& \frac{1}{2^4 \cdot 3 \pi}\int_{-\infty}^{\infty} d \omega~ \omega^2 S(\omega),\\
    \kappa^{(2)} &=& \frac{1}{2^5 \cdot 45\pi}\int_{-\infty}^{\infty} d \omega~\omega^4 S(\omega),
\end{eqnarray}
correspond to the integral over the power spectrum and its first two moments. Since one does not have access to $\{\kappa^{(0)},\kappa^{(1)},\kappa^{(2)}\}$ a priori, we employ a polynomial fitting procedure subject to the functional form in Eq.~\eqref{eq:chi-short-wt-expansion-FID} to obtain values for the attenuation function, $\chi_{\mathrm{FID}}(t)$, over the interval $[0,\tau]$. To ensure physically correct behavior for the interpolated $\chi_{\mathrm{FID}}(t)$ in the short-time region, we employ two additional fitting constraints: $C(t \rightarrow 0) = 1$ and in the region at and beyond $\tau$, the fitting procedure must align with the first few measured values. Thus, we perform the polynomial fitting in the interval $[0,\tau+\epsilon]$ where $\epsilon$ contains the first few points accessible via direct measurement of the coherence curve. This ensures that the inferred values of these constants are correctly reconstructing the expected coherence curve well into the $\epsilon$ interval that one can directly measure. We expect that, depending on the structure of the noise and the resulting coherence function, one might need to keep more terms in the expansion above to be able to infer the points in the $[0,\tau+\epsilon]$ interval in future applications.

We apply similar expansion to the spin echo sequence to reconstruct the early time behavior of the corresponding attenuation function. In this case,
\begin{equation} \label{eq:chi-short-wt-expansion-SE}
\begin{split}
    \chi_{\mathrm{SE}}(t) &= \frac{4}{\pi} \int_{-\infty}^{\infty} d \omega~ \frac{1}{\omega^2}S(\omega)\sin^4\Big( \frac{\omega t}{2}\Big)\\
    &\approx \frac{4}{\pi} \int_{-\infty}^{\infty} d \omega~ \frac{1}{\omega^2}S(\omega)\Big[ \frac{(\omega t)^4}{2^4}-\frac{(\omega t)^6}{2^6 \cdot \frac{3}{2}}+\frac{(\omega t)^8}{2^8 \cdot 5}\Big] \\ 
    &\equiv \kappa^{(0)} t^4+\kappa^{(1)} t^6+\kappa^{(2)} t^8.
\end{split}
\end{equation} 
Here, one can find the parameters via the same fitting procedure as described above. This early time reconstruction of the spin echo sequence both enables the implementation of the FTNS protocol and allows the spin echo DDNS to go beyond the spectral limit set by the minimum delay time. This can be seen in Figs.~5 and 6 where we remove the frequency limitation of the DDNS reconstruction set by the minimum delay time.

Furthermore, depending on the quality of the obtained fit, it may be necessary to modulate the transition between the fitted data and the measurement data if the resulting first and second numerical derivatives show large fluctuations at the boundary. This can be done by multiplying the measured data by a shifted Error function which has a transition length of about $5 \delta t$ and a transition point at about $5 \delta t + \tau_\mathrm{min}$, and also multiplying the fitted data by the negative of the same Error function. These two sets of data can now be added to give a modulated time series data with a suppressed effect of any discontinuities arising in the transition from the fitted data to the measured data. We employed this procedure in the early time reconstruction implemented in Fig.~2.

In all examples shown in this paper except for Fig.~1, we have restricted the total measurement time such that the coherence value does not become less than $C(t)=0.005$. This ensures that the measured values remain within the reasonable experimentally accessible range. 
%Furthermore, in all cases, we employed a minimum delay time such that $\tau/ \delta t= 16$ for both FTNS and DDNS. 

%%%%%%%%%%%%%%%%%%%%%%%%%%%%%%%%%%%%%%%%%%%%%%%%%%%%%%%%%%%%%%%%%%%%%%%%%%%
%%%%%%%%%%%%%%%%%%%%%%%%%%%%%%%%%%%%%%%%%%%%%%%%%%%%%%%%%%%%%%%%%%%%%%%%%%%
%%%%%%%%%%%%%%%%%%%%%%%%%%%%%%%%%%%%%%%%%%%%%%%%%%%%%%%%%%%%%%%%%%%%%%%%%%%
%%%%%%%%%%%%%%%%%%%%%%%%%%%%%%%%%%%%%%%%%%%%%%%%%%%%%%%%%%%%%%%%%%%%%%%%%%%

\subsection{\textbf{Linear behavior of $\chi(t)$ at long times}} \label{app-linear}

Here, we demonstrate that $\ddot{\chi}_{\mathrm{FID}}(t)\to 0$ at $t\to\infty$ {\it for any spectrum whose integral over all frequencies remains finite.} To do this, we recall the Riemann-Lebesgue lemma, which states that the Fourier transform of a function $f(\omega)$ vanishes as $t\rightarrow \infty$, as long as
\begin{equation}
    \int_{-\infty}^\infty \vert f(\omega) \vert\ d\omega < \infty.
\end{equation}
In FTNS, $\ddot{\chi}_\mathrm{FID}(t)$ is equal to the inverse Fourier transform of $S(\omega)/\sqrt{2 \pi}$. Thus, $\ddot{\chi}_{\mathrm{FID}}(t\to\infty)\to 0$ is simply a consequence of the Riemann-Lebesgue lemma so long as the noise spectrum $S(\omega)$ is of a functional form whose area under the curve is finite, which is a condition many physical noise spectra are expected to obey. This result guarantees that under such conditions, $\chi_\mathrm{FID}(t)$ can only grow at most linearly in $t$ at $t\to\infty$. This ensures that fitting $\chi_\mathrm{FID}(t)$ to a linear function at long $t$ is a valid method for mitigating the effects of measurement error for many physical systems.

For concreteness, we now explicitly show that $\chi_{\mathrm{FID}}(t)$ behaves linearly at $t\to\infty$ for the Lorentzian and the Gaussian spectra, which are two commonly encountered spectral shapes. To do this, we consider a generic form for $S(\omega)$ and obtain an expression for $\chi_{\mathrm{FID}}(t)$. The $t\to\infty$ behavior of this $\chi_{\mathrm{FID}}(t)$ reveals the expected linear behavior. {\it We reiterate that these are specific examples of a general result that holds for any realistic noise spectrum whose integral over all frequencies remains finite.}

%\subsubsection*{Lorentzian spectrum}

We first consider a Lorentzian spectrum:
\begin{equation}
    S(\omega)=A\left(\frac{1}{1+\left(\frac{\omega-d}{\omega_\mathrm{c}}\right)^2}+\frac{1}{1+\left(\frac{\omega+d}{\omega_\mathrm{c}}\right)^2}\right).
\end{equation}
This form ensures that it is symmetric. Taking the inverse Fourier transform of $S(\omega)/\sqrt{2\pi}$, we obtain,
\begin{equation}
    \ddot{\chi}_{\mathrm{FID}}(t) =  A \omega_\mathrm{c} \mathrm{e}^{-t \omega_\mathrm{c}} \cos (d t).
\end{equation}
Clearly, this is a function that decays exponentially to zero at long times. From this we can obtain $\dot{\chi}_{\mathrm{FID}}(t)$ and $\chi_{\mathrm{FID}}(t)$:
\begin{widetext}
\begin{eqnarray}
    \dot{\chi}_{\mathrm{FID}}(t) &=& \frac{  A \omega_\mathrm{c} \mathrm{e}^{-t \omega_\mathrm{c}} (d \sin (d t)-\omega_\mathrm{c} \cos (d t))}{d^2+\omega_\mathrm{c}^2} + C_1,\\
    \chi_{\mathrm{FID}}(t) &=& \frac{ A \omega_\mathrm{c} \mathrm{e}^{-t \omega_\mathrm{c}} \left(\left(\omega_\mathrm{c}^2-d^2\right) \cos (d t)-2 d \omega_\mathrm{c} \sin (d
   t)\right)}{\left(d^2+\omega_\mathrm{c}^2\right)^2} + C_1 t + C_2,
\end{eqnarray}
\end{widetext}

\noindent where $C_1$ and $C_2$ are integration constants, which we can find by enforcing the appropriate boundary conditions. The coherence should start at 1 at $t=0$, so we expect $\chi_{\mathrm{FID}}(t=0)=0$. We can also examine the boundary condition for $\dot{\chi}_{\mathrm{FID}}(t)$:
\begin{eqnarray}
\dot{\chi}_{\mathrm{FID}}(t) &=& \frac{1}{\pi} \int_{-\infty}^{\infty} d\omega\  \frac{S(\omega)}{\omega}  \sin (\frac{\omega t}{2})  \cos (\frac{\omega t}{2}),
\end{eqnarray}
which implies that $\dot{\chi}(t =0) = 0$. We impose these by evaluating $\dot{\chi}(0)$ and $\chi(0)$:
\begin{eqnarray}
\dot{\chi}_{\mathrm{FID}}(0) &=& -\frac{ A \omega_\mathrm{c}^2}{d^2+\omega_\mathrm{c}^2} + C_1 = 0, \\
\chi_{\mathrm{FID}}(0) &=& \frac{ A \omega_\mathrm{c} \left(\omega_\mathrm{c}^2-d^2\right)}{\left(d^2+\omega_\mathrm{c}^2\right)^2}  + C_2 = 0.
\end{eqnarray}
Hence,
\begin{widetext}
\begin{equation}
    \chi_{\mathrm{FID}}(t) = \frac{ A \omega_\mathrm{c} \mathrm{e}^{-t \omega_\mathrm{c}} \left(\left(\omega_\mathrm{c}^2-d^2\right) \cos (d t)-2 d \omega_\mathrm{c} \sin (d
   t)\right)}{\left(d^2+\omega_\mathrm{c}^2\right)^2} + \frac{ A \omega_\mathrm{c}^2}{d^2+\omega_\mathrm{c}^2} t - \frac{ A \omega_\mathrm{c} \left(\omega_\mathrm{c}^2-d^2\right)}{\left(d^2+\omega_\mathrm{c}^2\right)^2}.
\end{equation}
\end{widetext}

Thus, the long-time limit of the attenuation function is a linear function in $t$,
\begin{equation}
    \lim_{t\to\infty} \chi_{\mathrm{FID}}(t) = \frac{ A \omega_\mathrm{c}^2}{d^2+\omega_\mathrm{c}^2} t - \frac{ A \omega_\mathrm{c} \left(\omega_\mathrm{c}^2-d^2\right)}{\left(d^2+\omega_\mathrm{c}^2\right)^2}.
\end{equation}

%\subsubsection*{Gaussian spectrum}

One can perform a similar analysis for a Gaussian spectrum,
\begin{equation}
    S(\omega) = A \exp\left\{ -\left(\frac{\omega-d}{\sigma} \right)^2\right\} + A \exp\left\{ -\left(\frac{\omega+d}{\sigma} \right)^2\right\}.
\end{equation}

We can again take the inverse Fourier transform of $S(\omega)/\sqrt{2\pi}$ to obtain,
\begin{equation}\label{eq:ddchi-Gaussian}
    \ddot{\chi}_{\mathrm{FID}}(t) = \frac{A}{2\sqrt{\pi}} \vert \sigma \vert \exp\left\{ -\frac{1}{4} t (4 \mathrm{i} d+ t \sigma^2) \right\} \left( 1+\mathrm{e}^{2 \mathrm{i} d t} \right),
\end{equation}
which goes to zero at long times. We then integrate Eq.~\ref{eq:ddchi-Gaussian} to obtain expressions for $\dot{\chi}(t)$ and $\chi(t)$ subject to their constraints at $t \rightarrow 0$, i.e., $\chi(0) = 0$ and $\dot{\chi}(0) = 0$:
\begin{widetext}
\begin{eqnarray}
    \dot{\chi}_{\mathrm{FID}}(t) &=&  A \mathrm{e}^{-\frac{d^2}{\sigma ^2}} \Re\left(\text{Erf}\left(\frac{\mathrm{i} d}{\sigma }+\frac{\sigma  t}{2}\right)\right),  \\
   \chi_{\mathrm{FID}}(t) &=& -\frac{2 A \left(\sigma  +\mathrm{i} \mathrm{e}^{-\frac{d^2}{\sigma ^2}} \sqrt{\pi } d ~\text{Erf}\left( \frac{\mathrm{i} d}{\sigma
   }\right)\right)}{\sqrt{\pi} \sigma ^2}\nonumber \\
   &&+ \Re\left(\frac{A \mathrm{e}^{-\frac{d^2}{\sigma ^2}} \left(2 \sigma  \mathrm{e}^{\left(\frac{d}{\sigma }+\frac{\mathrm{i} \sigma 
   t}{2}\right)^2}+\mathrm{i}\sqrt{\pi } \left(2 d+\mathrm{i} \sigma ^2 t\right) ~\text{Erf}\left(\frac{\mathrm{i}d}{\sigma }-\frac{ \sigma  t}{2}\right)\right)}{\sqrt{\pi} \sigma ^2}\right),
\end{eqnarray}
\end{widetext}

\noindent where $\Re(\cdot)$ denotes the real part, and Erf$(\cdot)$ is the error function. Noting that $\lim_{t \rightarrow \infty} \mathrm{Erf}(t) \to 1$, it is clear that the long-time limit of the attenuation function becomes,
\begin{equation}
    \lim_{t\to\infty} \chi_{\mathrm{FID}}(t) = \frac{ A}{\sigma^2} \left(\mathrm{e}^{-\frac{d^2}{\sigma ^2}} \left(-2 \mathrm{i} d ~\text{Erf}\left(\frac{\mathrm{i} d}{\sigma }\right)+\sigma ^2 t\right)-\frac{2 \sigma}{\sqrt{\pi }}
   \right).
\end{equation}
Hence, the long-time behavior of $\chi_{\mathrm{FID}}(t)$ for a Gaussian-shaped power spectrum is also linear in $t$, as expected from the Riemann-Lebesgue lemma.

%%%%%%%%%%%%%%%%%%%%%%%%%%%%%%%%
\begin{figure*}
\hspace{-.2cm}{\includegraphics[scale=.07]{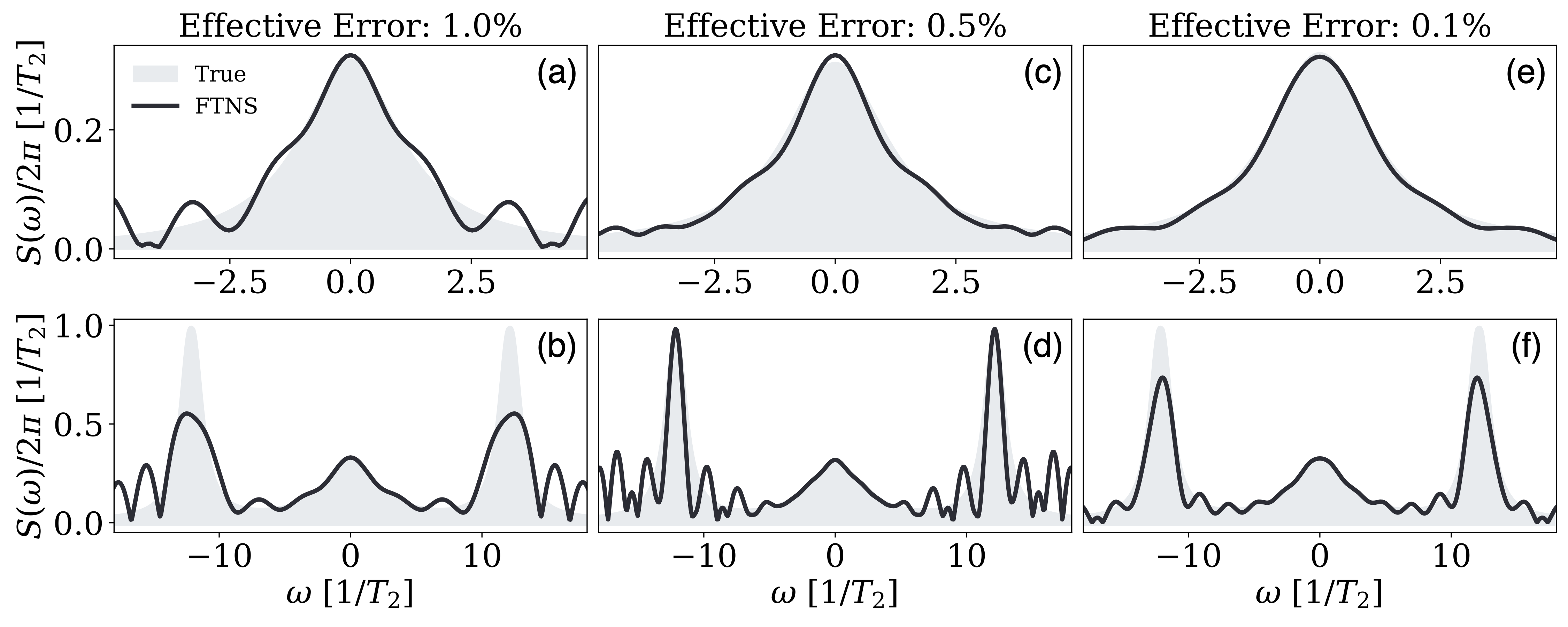}}
\caption{\textbf{Noise spectrum reconstructions using the FTNS method under different values of effective simulated measurement noise}: FTNS in the presence of various effective measurement errors, for the two examples demonstrated in Fig.~4. Specifically, the noise levels for the panels are $1.0\%$ for (a) and (b), $0.5\%$ for  (c) and (d), and $0.1\%$ for (e) and (f). While performance improves for lower effective noise, the peaked spectral features remain more or less robust under all three cases.}
\label{fig-sm}
\end{figure*}
%%%%%%%%%%%%%%%%%%%%%%%%%%%%%%%%

%%%%%%%%%%%%%%%%%%%%%%%%%%%%%%%%%%%%%%%%%%%%%%%%%%%%%%%%%%%%%%%%%%%%%%%%%%%
%%%%%%%%%%%%%%%%%%%%%%%%%%%%%%%%%%%%%%%%%%%%%%%%%%%%%%%%%%%%%%%%%%%%%%%%%%%
%%%%%%%%%%%%%%%%%%%%%%%%%%%%%%%%%%%%%%%%%%%%%%%%%%%%%%%%%%%%%%%%%%%%%%%%%%%
%%%%%%%%%%%%%%%%%%%%%%%%%%%%%%%%%%%%%%%%%%%%%%%%%%%%%%%%%%%%%%%%%%%%%%%%%%%
\subsection{\textbf{Error mitigation protocol}} \label{app-spam} 

Here we outline the details of our approach to mitigate measurement errors. In particular, we detail the protocol we developed and employed to generate Fig.~4 from noisy coherence measurements. As a demonstration, we have used \textit{Mathematica}, but our protocol is general and can be implemented within other computational softwares. We emphasize that this is \textit{one example} of a denoising protocol; other procedures may be more appropriate for different data and physical problems. 

In Fig.~4, we model the noise in the coherence function as arising from a normal distribution with mean 0 and standard deviation 0.001 at each measurement point. The noise has been adjusted such that for early times, the acquired value for $C(t)$ does not exceed unity, and at later times it does not fall below zero.

We now summarize our denoising protocol: 
\begin{enumerate}
    \item Mirror the coherence data around $t = 0$ to get an effective coherence profile from $-T_\mathrm{max}$ to $T_\mathrm{max}$. This allows the numerical time-derivative to obtain a better value of $\chi(t)$ at $t = 0$, which helps to improve the performance of the Fourier transform near $\omega=0$.
    
    \item Process the noisy coherence data through a low-pass filter, with the cutoff frequency set to half of the sampling rate. All instances of the low-pass filter are implemented using the Mathematica built-in \texttt{LowpassFilter}. 
    
    \item Take the logarithm of the smoothed coherence to get effective $\chi(t)$ values.
    
    \item Plot the resulting data to visually discern whether the late time behavior appears linear and within what range a linear fit appears suitable. In this case, we determined that linear fits from $t/T_\text{max}=0.577$ to $t/T_\text{max}=0.990$ for Fig.~4(a) and the $0.5\%$ and $0.1\%$ effective error panels for the Gaussian spectrum in Fig.~7, and from $t/T_\text{max}=0.539$ to $t/T_\text{max}=0.987$ for Fig.~4(b) and the $0.5\%$ and $0.1\%$ effective error panels for the Lorentzian spectrum in Fig.~7, were appropriate. The justification for this linear fitting at long times is given in Sec.~II of this SM. For the linear fitting in the $1.0\%$ effective error panels, we used $t/T_\mathrm{max}$ ranges [0.577, 0.825] for the Gaussian spectrum and [0.359, 0.718] for the Lorentzian spectrum.
    
    \item Perform a linear fit on the ranges selected. We employed the \texttt{Fit} function in Mathematica.
    
    \item Replace the data within the selected range with the linear fit. This leads to a modified $\chi(t)$, which we denote by $\tilde{\chi}(t)$.
    
    \item \textit{Optional:} After applying a linear fit, one can extend the $\tilde{\chi}(t)$ data to arbitrarily long times, which results in a longer effective measurement time, which in turn provides improved resolution in frequency space of the FTNS approach. This step was not implemented in the generation of Fig.~4 in the main text and its implementation would only increase the frequency resolution of the spectrum.
    
    \item Perform a numerical time derivative of $\tilde{\chi}(t)$. To obtain the numerical time derivatives, we implemented first order forward and backward difference approximation on the first and last data points, and a second order centered-difference approximation on the rest of the points. This is the algorithm behind various differentiation packages, such as \texttt{numpy.gradient}, which we used for the simulations in Figs.~1, 2, and 3 in the main text.
    
    \item If the linear fitting causes a discontinuity, we remove its effect on the derivative by setting the value of the first derivative at the discontinuity to the derivative of the linear fit. 
    
    \item Apply another low-pass filter at a cutoff frequency at 1/4 of the sampling rate. Note that this step was implemented for all Gaussian spectra in Figs.~4 and 7, but not for the Lorentzian spectra in Figs.~4 and 7.
    
    \item We take a second numerical time-derivative of the data.
    
    \item We apply a Fourier transform on the data as discussed in the main text: one can, for example, use any FFT implementation available in numerical packages (e.g., \texttt{numpy}) or implement the Fourier transform manually by performing an integral of the quantity $\ddot{\tilde{\chi}}(t) \mathrm{e}^{\mathrm{i} \omega t}/\sqrt{2 \pi}$ over time, where the integration is approximated by the trapezoidal rule without changing the result. 
    
    \item For the Fourier transform, we employed a frequency range from $\pm$ half of the sampling rate of the coherence, with $\delta t/T_\mathrm{max} =0.002$ for Figs.~4(a) and (b). Finally, this is divided by $\sqrt{2 \pi}$ to obtain the denoised spectra seen in Fig.~4.
    
\end{enumerate}

We can study the performance of FTNS using this particular denoising protocol at various effective measurement error percentages. Figure 7 gives examples of this for the two spectra used Fig.~4 in the main text at effective noise values of $1.0\%$, $0.5\%$, and $0.1\%$. As expected, lower noise values give better agreement with the true spectrum. Yet, the agreement between FTNS and the true spectrum demonstrates that FTNS can robustly capture the major peaks in the spectrum in all cases. Strikingly, the artifacts of the Fourier transform of noisy data, which are most prominent in the examples with $1.0\%$ noise levels, systematically decrease with increased sampling. Thus, to robustly identify features of the true spectrum in an experimental implementation of FTNS, it would be helpful to compare averages of smaller batches of measurements for common peaked features that appear consistently. Such comparisons can also be used to check the convergence of the reconstructed spectrum as a function of the extent of averaging done during the measurement process. 

\subsection{\textbf{Noise spectroscopy protocol for $1/f$-type spectra}} \label{app-SE-1overf} 

Here, we show the long-time behavior of $\chi_\mathrm{SE}(t)$ under both $1/f$ and integrable spectra, and use these results to formulate a noise spectroscopy protocol for spectra consisting of both $1/f$ and finite peaked features.

%\subsubsection*{Long-time behavior of $\chi_\mathrm{SE}(t)$ under $1/f$ noise}
First we show the long-time behavior of $\chi_\mathrm{SE}(t)$ under $1/f$ noise. From Eq.~\eqref{eq-SE-FTNS},
\begin{equation}
\mathcal{F}[\ddot{\chi}_{\mathrm{SE}}(t)]= \sqrt{\frac{2}{\pi}} M(2\omega) = \sqrt{\frac{2}{\pi}} \left( S(2\omega)-S(\omega)/2 \right) \nonumber,
\end{equation}
we can derive the analytic form of $\chi_\mathrm{SE}(t)$ expected for $1/f$-type spectra in the long time limit, in a similar manner to what was shown in Methods section~\ref{app-linear}. 

For 
\begin{equation}
S_\mathrm{1/f}(\omega) = \mathcal{A}/\omega^n \label{eq:Sw_1overf}
\end{equation}
with the constraint that $0< n < 3$, we obtain for $\chi(t)$ using spin echo FTNS
\begin{equation}
\chi_\mathrm{SE,1/f}(t) = \mathcal{Y}_n t^{n+1},\label{eq:chi_1overf}
\end{equation}
where $\mathcal{Y}_n$ is given in Eq.~\eqref{eq-y_n}. Note the characteristic $t^{n+1}$ dependence which, under our assumption that $n>0$, grows faster than linear in $t$.

These results allow us to perform noise spectroscopy on spectra consisting of a linear combination of such functions, and therefore on generic $\propto 1/\omega^n$ type spectra.

%\subsubsection*{Long-time behavior of $\chi_\mathrm{SE}(t)$ under integrable noise spectra}
We now study the long-time behavior of $\chi_\mathrm{SE}(t)$ under integrable noise spectra.
For generic noise spectra which satisfy
\begin{equation}
\int_{-\infty}^\infty \vert M(2\omega) \vert d\omega < \infty, \label{eq:SE-longtime-lin-condition}
\end{equation}
the result 
\begin{equation}
    \ddot{\chi}_\mathrm{SE}(t\to\infty) \to 0
\end{equation}
holds as a consequence of the Riemann-Lebesgue lemma (see Methods section~\ref{app-linear}). This means that $\chi_\mathrm{SE}(t\to\infty) \propto t$ should hold as long as Eq.~\ref{eq:SE-longtime-lin-condition} is satisfied. Note that this condition is explicitly violated for a $1/f$-type noise and so $1/f$ noise does not exhibit linear $t$ dependence at $t\to\infty$.

%\subsubsection*{Noise spectroscopy of $1/f$ noise} \label{app-1overf}

The above results indicate that if we assume the total noise spectrum to be a linear combination of instances of spectra satisfying either category, we may be able to isolate the $1/f$ contribution through a fit of the measured $\chi(t)$ to a function of the form
\begin{equation}
\chi_\mathrm{fit}(t) = \alpha \vert t\vert^\gamma + \beta t + \delta. \label{eq:chi_fit}
\end{equation}
Eq.~\ref{eq:chi_1overf} shows that the parameters $\alpha$ and $\gamma$ fully characterize the $1/f$ component. Knowing these parameters, one can subtract off the contribution due to the $1/f$ component to obtain any residual structure in the noise spectrum. That is, if
\begin{equation}
\chi(t) = \chi_\mathrm{1/f}(t)+\chi_\mathrm{residual}(t),
\end{equation}
one can fit $\chi(t)$ to Eq.~\ref{eq:chi_fit} to obtain values for $\alpha$ and $\gamma$, which can then be used to find the parameters characterizing the $1/f$ noise, $\mathcal{A}$ and $n$, like
\begin{eqnarray}
n &=& \gamma-1,\label{eq:1overf_param_n} \\
\mathcal{A} &=&-\frac{ 2^{\gamma } \pi \alpha \csc \left(\frac{1}{2} \pi  (\gamma -1)\right)}{\left(2^{\gamma }-4\right) \Gamma (-\gamma )},\label{eq:1overf_param_A}
\end{eqnarray}
from Eqs.~\ref{eq:Sw_1overf} and \ref{eq:chi_1overf}. One can then use these parameters to reconstruct $\chi_\mathrm{1/f}$ at the time points corresponding to those obtained in the measurement $\chi(t)$. A pointwise difference between the measured $\chi(t)$ data and the reconstructed $\chi_\mathrm{1/f}(t)$ data yields an effective $\chi_\mathrm{residual}(t)$. After removing the $1/f$ component, the residual response should only have a linear dependence in time at long times. At this point, one can apply the iterative spin echo FTNS procedure outlined in the ``Spin-Echo FTNS'' section in the main text to reveal any additional structure in the noise spectrum.

For the specific examples shown in Fig.~6, we employed the following procedure:

\begin{enumerate}
\item Extract $\chi(t)$ from the measured $C(t)$ using Eq.~\ref{eq-coherence-def}. In our numerical example in Fig.~6, we used $S(\omega)$ given by 
\begin{equation} \label{eq:1overf_Sw_methods}
    S(\omega) = \frac{\mathcal{A}}{\omega^n} + \frac{B}{1+\left(\frac{\omega-d}{\omega_\mathrm{c}}\right)^2}+\frac{B}{1+\left(\frac{\omega+d}{\omega_\mathrm{c}}\right)^2},
\end{equation}
with example parameter values $\mathcal{A}=1, \,\,B=1, \,\,n=2.5, \,\,\omega_\mathrm{c} = 1.5, \,\,d=12.5$, and the spin echo filter function. In our demonstration, the experimentally motivated constraint that only measurements up to a final time such that all $C(t) < 0.005$ should be taken was implemented.
\item We then fit this measured $\chi(t)$ data to a function of the form given by Eq.~\ref{eq:chi_fit}. In our demonstration, we employed the {\tt FindFit} function in Mathematica to obtain parameter values
\begin{eqnarray}
&&\alpha = 0.0385433, \,\, \beta = 0.0223275, \nonumber \\ 
&&\delta = 0.0167974, \,\, \gamma = 3.51096. \nonumber
\end{eqnarray}
\item We use the $\alpha$ and $\gamma$ values obtained from the fit to construct the $1/f$ part of the spectrum via Eqs.~\ref{eq:1overf_param_n} and \ref{eq:1overf_param_A}. For our demonstration, we obtain
\begin{equation}
    S_{1/f}(\omega) = 0.974526/\omega^{2.51095}.
\end{equation}
Note that the effective $\mathcal{A}$ and $n$ values obtained from the fit are in reasonable agreement with their exact values, $\mathcal{A}=1$ and $n=2.5$.
\item From the effective $\mathcal{A}$ and $n$ values obtained from the previous step, we generate a time series data of $\chi_\mathrm{1/f}(t)$ at time points corresponding to the measurement times of the original data, via Eq.~\ref{eq:chi_1overf}. We then subtract these values, pointwise, from the original measurement data $\chi(t)$ to obtain $\chi_\mathrm{residual}(t)$. $\chi_\mathrm{residual}(t)$ corresponds to the component of the attenuation function that is due to all except the $1/f$ component of the noise spectrum.
\item We then perform the spin echo FTNS protocol, as outlined in the main text, on $\chi_\mathrm{residual}(t)$. This provides an additional contribution to the total noise spectrum, $S_\mathrm{residual}(\omega)$.
\item Finally, we add the resulting $S_\mathrm{1/f}(\omega)$ and $S_\mathrm{residual}(\omega)$ to obtain the total reconstructed spectrum. We show in Fig.~6 of the main text that this noise spectroscopy protocol can sufficiently characterize the nature of the $1/f$ approach of the spectrum at $\omega \rightarrow 0$, as well as identify the presence of any additional high-frequency peaks which may be present in the noise spectrum.
\end{enumerate}

\section*{DATA AVAILABILITY}

The data that support the findings of this study are available from the authors upon request.

\section*{CODE AVAILABILITY}
The code used to generate the results of this study are available from the authors upon request.

\begin{acknowledgments}

A.M.C and S.S.~acknowledges funding from National Science Foundation (Grant No. 2326837). A.M.C.~acknowledges the start-up funds from the University of Colorado Boulder. S.S.~acknowledges funding from National Science Foundation (Grant No. 2016244 and 2317149) and the Sloan Research Fellowship.
\end{acknowledgments}

\section*{COMPETING INTERESTS}
The authors declare that there are no competing interests.

\section*{AUTHOR CONTRIBUTIONS}
A.M.C. designed and supervised the project. A.V. and N.S. performed the analysis and numerical simulations. S.S. provided input for experimental feasibility. All authors contributed to the writing of the manuscript. A.V. and N.S. have contributed equally to this work.


\begin{thebibliography}{10}
\expandafter\ifx\csname url\endcsname\relax
  \def\url#1{\texttt{#1}}\fi
\expandafter\ifx\csname urlprefix\endcsname\relax\def\urlprefix{URL }\fi
\providecommand{\bibinfo}[2]{#2}
\providecommand{\eprint}[2][]{\url{#2}}

\bibitem{Cywinski2008PRB}
\bibinfo{author}{Cywi{\'n}ski, {\L}.}, \bibinfo{author}{Lutchyn, R.~M.},
  \bibinfo{author}{Nave, C.~P.} \& \bibinfo{author}{Sarma, S.~D.}
\newblock \bibinfo{title}{How to enhance dephasing time in superconducting
  qubits}.
\newblock \emph{\bibinfo{journal}{Phys. Rev. B}} \textbf{\bibinfo{volume}{77}},
  \bibinfo{pages}{174509} (\bibinfo{year}{2008}).

\bibitem{Biercuk2011JP}
\bibinfo{author}{Biercuk, M.~J.}, \bibinfo{author}{Doherty, A.~C.} \&
  \bibinfo{author}{Uys, H.}
\newblock \bibinfo{title}{Dynamical decoupling sequence construction as a
  filter-design problem}.
\newblock \emph{\bibinfo{journal}{J. Phys. B: At. Mol. Opt. Phys.}}
  \textbf{\bibinfo{volume}{44}}, \bibinfo{pages}{154002}
  (\bibinfo{year}{2011}).

\bibitem{Uhrig2007PRL}
\bibinfo{author}{Uhrig, G.~S.}
\newblock \bibinfo{title}{Keeping a quantum bit alive by optimized
  $\ensuremath{\pi}$-pulse sequences}.
\newblock \emph{\bibinfo{journal}{Phys. Rev. Lett.}}
  \textbf{\bibinfo{volume}{98}}, \bibinfo{pages}{100504}
  (\bibinfo{year}{2007}).

\bibitem{Biercuk2009PRA}
\bibinfo{author}{Biercuk, M.~J.} \emph{et~al.}
\newblock \bibinfo{title}{Experimental uhrig dynamical decoupling using trapped
  ions}.
\newblock \emph{\bibinfo{journal}{Phys. Rev. A}} \textbf{\bibinfo{volume}{79}},
  \bibinfo{pages}{062324} (\bibinfo{year}{2009}).

\bibitem{Yang2010Fphys}
\bibinfo{author}{Yang, W.}, \bibinfo{author}{Wang, Z.-Y.} \&
  \bibinfo{author}{Liu, R.-B.}
\newblock \bibinfo{title}{Preserving qubit coherence by dynamical decoupling}.
\newblock \emph{\bibinfo{journal}{Front. Phys. China}}
  \textbf{\bibinfo{volume}{6}}, \bibinfo{pages}{2--14} (\bibinfo{year}{2010}).

\bibitem{Szakowski2017JP}
\bibinfo{author}{Sza{\'{n}}kowski, P.}, \bibinfo{author}{Ramon, G.},
  \bibinfo{author}{Krzywda, J.}, \bibinfo{author}{Kwiatkowski, D.} \&
  \bibinfo{author}{Cywi{\'{n}}ski, {\L}.}
\newblock \bibinfo{title}{Environmental noise spectroscopy with qubits
  subjected to dynamical decoupling}.
\newblock \emph{\bibinfo{journal}{J. Phys.: Condens. Matter}}
  \textbf{\bibinfo{volume}{29}}, \bibinfo{pages}{333001}
  (\bibinfo{year}{2017}).

\bibitem{DegenRevModPhys2017}
\bibinfo{author}{Degen, C.~L.}, \bibinfo{author}{Reinhard, F.} \&
  \bibinfo{author}{Cappellaro, P.}
\newblock \bibinfo{title}{Quantum sensing}.
\newblock \emph{\bibinfo{journal}{Rev. Mod. Phys.}}
  \textbf{\bibinfo{volume}{89}}, \bibinfo{pages}{035002}
  (\bibinfo{year}{2017}).

\bibitem{Suter2016Review}
\bibinfo{author}{Suter, D.} \& \bibinfo{author}{\'Alvarez, G.~A.}
\newblock \bibinfo{title}{Colloquium: Protecting quantum information against
  environmental noise}.
\newblock \emph{\bibinfo{journal}{Rev. Mod. Phys.}}
  \textbf{\bibinfo{volume}{88}}, \bibinfo{pages}{041001}
  (\bibinfo{year}{2016}).

\bibitem{Alvarez2011PRL}
\bibinfo{author}{\'Alvarez, G.~A.} \& \bibinfo{author}{Suter, D.}
\newblock \bibinfo{title}{Measuring the spectrum of colored noise by dynamical
  decoupling}.
\newblock \emph{\bibinfo{journal}{Phys. Rev. Lett.}}
  \textbf{\bibinfo{volume}{107}}, \bibinfo{pages}{230501}
  (\bibinfo{year}{2011}).

\bibitem{Yuge2011PRL}
\bibinfo{author}{Yuge, T.}, \bibinfo{author}{Sasaki, S.} \&
  \bibinfo{author}{Hirayama, Y.}
\newblock \bibinfo{title}{Measurement of the noise spectrum using a
  multiple-pulse sequence}.
\newblock \emph{\bibinfo{journal}{Phys. Rev. Lett.}}
  \textbf{\bibinfo{volume}{107}}, \bibinfo{pages}{170504}
  (\bibinfo{year}{2011}).

\bibitem{Norris2016PRL}
\bibinfo{author}{Norris, L.~M.}, \bibinfo{author}{Paz-Silva, G.~A.} \&
  \bibinfo{author}{Viola, L.}
\newblock \bibinfo{title}{Qubit noise spectroscopy for non-gaussian dephasing
  environments}.
\newblock \emph{\bibinfo{journal}{Phys. Rev. Lett.}}
  \textbf{\bibinfo{volume}{116}}, \bibinfo{pages}{150503}
  (\bibinfo{year}{2016}).

\bibitem{Krzywda2019NJP}
\bibinfo{author}{Krzywda, J.}, \bibinfo{author}{Sza{\'{n}}kowski, P.} \&
  \bibinfo{author}{Cywi{\'{n}}ski, {\L}.}
\newblock \bibinfo{title}{The dynamical-decoupling-based spatiotemporal noise
  spectroscopy}.
\newblock \emph{\bibinfo{journal}{New J. Phys.}} \textbf{\bibinfo{volume}{21}},
  \bibinfo{pages}{043034} (\bibinfo{year}{2019}).

\bibitem{Bylander2011NatPhys}
\bibinfo{author}{Bylander, J.} \emph{et~al.}
\newblock \bibinfo{title}{Noise spectroscopy through dynamical decoupling with
  a superconducting flux qubit}.
\newblock \emph{\bibinfo{journal}{Nat. Phys.}} \textbf{\bibinfo{volume}{7}},
  \bibinfo{pages}{565--570} (\bibinfo{year}{2011}).

\bibitem{SungNature2019}
\bibinfo{author}{Sung, Y.} \emph{et~al.}
\newblock \bibinfo{title}{Non-gaussian noise spectroscopy with a
  superconducting qubit sensor}.
\newblock \emph{\bibinfo{journal}{Nat. Commun.}} \textbf{\bibinfo{volume}{10}},
  \bibinfo{pages}{3715} (\bibinfo{year}{2019}).

\bibitem{Almog2011JP}
\bibinfo{author}{Almog, I.} \emph{et~al.}
\newblock \bibinfo{title}{Direct measurement of the
  system{\textendash}environment coupling as a tool for understanding
  decoherence and dynamical decoupling}.
\newblock \emph{\bibinfo{journal}{J. Phys. B: At. Mol. Opt. Phys.}}
  \textbf{\bibinfo{volume}{44}}, \bibinfo{pages}{154006}
  (\bibinfo{year}{2011}).

\bibitem{Dial2013PRL}
\bibinfo{author}{Dial, O.~E.} \emph{et~al.}
\newblock \bibinfo{title}{Charge noise spectroscopy using coherent exchange
  oscillations in a singlet-triplet qubit}.
\newblock \emph{\bibinfo{journal}{Phys. Rev. Lett.}}
  \textbf{\bibinfo{volume}{110}}, \bibinfo{pages}{146804}
  (\bibinfo{year}{2013}).

\bibitem{farfurnik2023all}
\bibinfo{author}{Farfurnik, D.} \emph{et~al.}
\newblock \bibinfo{title}{All-optical noise spectroscopy of a solid-state
  spin}.
\newblock \emph{\bibinfo{journal}{Nano Lett.}} \textbf{\bibinfo{volume}{23}},
  \bibinfo{pages}{1781--1786} (\bibinfo{year}{2023}).

\bibitem{Connors2022NatCom}
\bibinfo{author}{Connors, E.}, \bibinfo{author}{Nelson, J.},
  \bibinfo{author}{Edge, L.} \& \bibinfo{author}{Nichol, J.}
\newblock \bibinfo{title}{Charge-noise spectroscopy of si/sige quantum dots via
  dynamically-decoupled exchange oscillations}.
\newblock \emph{\bibinfo{journal}{Nat. Commun.}} \textbf{\bibinfo{volume}{13}},
  \bibinfo{pages}{940} (\bibinfo{year}{2022}).

\bibitem{BarGill2012NatComm}
\bibinfo{author}{Bar-Gill, N.} \emph{et~al.}
\newblock \bibinfo{title}{Suppression of spin-bath dynamics for improved
  coherence of multi-spin-qubit systems}.
\newblock \emph{\bibinfo{journal}{Nat. Commun.}} \textbf{\bibinfo{volume}{3}},
  \bibinfo{pages}{858} (\bibinfo{year}{2012}).

\bibitem{Romach2015PRL}
\bibinfo{author}{Romach, Y.} \emph{et~al.}
\newblock \bibinfo{title}{Spectroscopy of surface-induced noise using shallow
  spins in diamond}.
\newblock \emph{\bibinfo{journal}{Phys. Rev. Lett.}}
  \textbf{\bibinfo{volume}{114}}, \bibinfo{pages}{017601}
  (\bibinfo{year}{2015}).

\bibitem{Carr1954}
\bibinfo{author}{Carr, H.~Y.} \& \bibinfo{author}{Purcell, E.~M.}
\newblock \bibinfo{title}{Effects of diffusion on free precession in nuclear
  magnetic resonance experiments}.
\newblock \emph{\bibinfo{journal}{Phys. Rev.}} \textbf{\bibinfo{volume}{94}},
  \bibinfo{pages}{630--638} (\bibinfo{year}{1954}).

\bibitem{Meiboom1958}
\bibinfo{author}{Meiboom, S.} \& \bibinfo{author}{Gill, D.}
\newblock \bibinfo{title}{Modified spin‐echo method for measuring nuclear
  relaxation times}.
\newblock \emph{\bibinfo{journal}{Rev. Sci. Instrum.}}
  \textbf{\bibinfo{volume}{29}}, \bibinfo{pages}{688--691}
  (\bibinfo{year}{1958}).

\bibitem{Szankowski2018PRA}
\bibinfo{author}{Sza{\'n}kowski, P.} \& \bibinfo{author}{Cywi{\'n}ski, {\L}.}
\newblock \bibinfo{title}{Accuracy of dynamical-decoupling-based spectroscopy
  of gaussian noise}.
\newblock \emph{\bibinfo{journal}{Phys. Rev. A}} \textbf{\bibinfo{volume}{97}},
  \bibinfo{pages}{032101} (\bibinfo{year}{2018}).

\bibitem{Wise2021PRX}
\bibinfo{author}{Wise, D.~F.}, \bibinfo{author}{Morton, J. J.~L.} \&
  \bibinfo{author}{Dhomkar, S.}
\newblock \bibinfo{title}{Using deep learning to understand and mitigate the
  qubit noise environment}.
\newblock \emph{\bibinfo{journal}{PRX Quantum}} \textbf{\bibinfo{volume}{2}},
  \bibinfo{pages}{010316} (\bibinfo{year}{2021}).

\bibitem{Yan2012PRB}
\bibinfo{author}{Yan, F.} \emph{et~al.}
\newblock \bibinfo{title}{Spectroscopy of low-frequency noise and its
  temperature dependence in a superconducting qubit}.
\newblock \emph{\bibinfo{journal}{Phys. Rev. B}} \textbf{\bibinfo{volume}{85}},
  \bibinfo{pages}{174521} (\bibinfo{year}{2012}).

\bibitem{Boss2016PRL}
\bibinfo{author}{Boss, J.~M.} \emph{et~al.}
\newblock \bibinfo{title}{One- and two-dimensional nuclear magnetic resonance
  spectroscopy with a diamond quantum sensor}.
\newblock \emph{\bibinfo{journal}{Phys. Rev. Lett.}}
  \textbf{\bibinfo{volume}{116}}, \bibinfo{pages}{197601}
  (\bibinfo{year}{2016}).

\bibitem{Gu2019}
\bibinfo{author}{Gu, B.} \& \bibinfo{author}{Franco, I.}
\newblock \bibinfo{title}{When can quantum decoherence be mimicked by classical
  noise?}
\newblock \emph{\bibinfo{journal}{J. Chem. Phys.}}
  \textbf{\bibinfo{volume}{151}}, \bibinfo{pages}{014109}
  (\bibinfo{year}{2019}).

\bibitem{PazSilva2017PRA}
\bibinfo{author}{Paz-Silva, G.~A.}, \bibinfo{author}{Norris, L.~M.} \&
  \bibinfo{author}{Viola, L.}
\newblock \bibinfo{title}{Multiqubit spectroscopy of gaussian quantum noise}.
\newblock \emph{\bibinfo{journal}{Phys. Rev. A}} \textbf{\bibinfo{volume}{95}},
  \bibinfo{pages}{022121} (\bibinfo{year}{2017}).

\bibitem{Kwiatkowski2020PRB}
\bibinfo{author}{Kwiatkowski, D.}, \bibinfo{author}{Sza{\'n}kowski, P.} \&
  \bibinfo{author}{Cywi{\'n}ski, {\L}.}
\newblock \bibinfo{title}{Influence of nuclear spin polarization on the
  spin-echo signal of an nv-center qubit}.
\newblock \emph{\bibinfo{journal}{Phys. Rev. B}}
  \textbf{\bibinfo{volume}{101}}, \bibinfo{pages}{155412}
  (\bibinfo{year}{2020}).

\bibitem{Mukamel1985}
\bibinfo{author}{Mukamel, S.}
\newblock \bibinfo{title}{Fluorescence and absorption of large anharmonic
  molecules - spectroscopy without eigenstates}.
\newblock \emph{\bibinfo{journal}{J. Phys. Chem.}}
  \textbf{\bibinfo{volume}{89}}, \bibinfo{pages}{1077--1087}
  (\bibinfo{year}{1985}).

\bibitem{mukamel1995}
\bibinfo{author}{Mukamel, S.}
\newblock \emph{\bibinfo{title}{Principles of Nonlinear Optical Spectroscopy}},
  vol.~\bibinfo{volume}{6} of \emph{\bibinfo{series}{Oxford Series in Optical
  and Imaging Sciences}} (\bibinfo{publisher}{Oxford University Press},
  \bibinfo{year}{1995}).

\bibitem{gradshteyn2014table}
\bibinfo{author}{Gradshteyn, I.~S.} \& \bibinfo{author}{Ryzhik, I.~M.}
\newblock \emph{\bibinfo{title}{Table of integrals, series, and products}}
  (\bibinfo{publisher}{Academic press}, \bibinfo{year}{2014}).

\bibitem{Makhlin2003JETPL}
\bibinfo{author}{Makhlin, Y.} \& \bibinfo{author}{Shnirman, A.}
\newblock \bibinfo{title}{Dephasing of qubits by transverse low-frequency
  noise}.
\newblock \emph{\bibinfo{journal}{JETP Lett.}} \textbf{\bibinfo{volume}{78}},
  \bibinfo{pages}{497--501} (\bibinfo{year}{2003}).

\bibitem{Schriefl2006NJP}
\bibinfo{author}{Schriefl, J.}, \bibinfo{author}{Makhlin, Y.},
  \bibinfo{author}{Shnirman, A.} \& \bibinfo{author}{Schön, G.}
\newblock \bibinfo{title}{Decoherence from ensembles of two-level fluctuators}.
\newblock \emph{\bibinfo{journal}{New J. Phys.}} \textbf{\bibinfo{volume}{8}},
  \bibinfo{pages}{1} (\bibinfo{year}{2006}).

\bibitem{Tyryshkin2010}
\bibinfo{author}{Tyryshkin, A.~M.} \emph{et~al.}
\newblock \bibinfo{title}{Dynamical decoupling in the presence of realistic
  pulse errors} (\bibinfo{year}{2010}).
\newblock \urlprefix\url{https://arxiv.org/abs/1011.1903}.

\bibitem{Souza2011PRL}
\bibinfo{author}{Souza, A.~M.}, \bibinfo{author}{\'Alvarez, G.~A.} \&
  \bibinfo{author}{Suter, D.}
\newblock \bibinfo{title}{Robust dynamical decoupling for quantum computing and
  quantum memory}.
\newblock \emph{\bibinfo{journal}{Phys. Rev. Lett.}}
  \textbf{\bibinfo{volume}{106}}, \bibinfo{pages}{240501}
  (\bibinfo{year}{2011}).

\bibitem{Chartrand2011IAM}
\bibinfo{author}{Chartrand, R.}
\newblock \bibinfo{title}{Numerical differentiation of noisy, nonsmooth data}.
\newblock \emph{\bibinfo{journal}{Int. Sch. Res. Notices}}
  \textbf{\bibinfo{volume}{2011}}, \bibinfo{pages}{164564}
  (\bibinfo{year}{2011}).

\bibitem{VanBreugel2020ieee}
\bibinfo{author}{Van~Breugel, F.}, \bibinfo{author}{Kutz, J.~N.} \&
  \bibinfo{author}{Brunton, B.~W.}
\newblock \bibinfo{title}{Numerical differentiation of noisy data: A unifying
  multi-objective optimization framework}.
\newblock \emph{\bibinfo{journal}{IEEE Access}} \textbf{\bibinfo{volume}{8}},
  \bibinfo{pages}{196865--196877} (\bibinfo{year}{2020}).

\bibitem{Press2010NatPhot}
\bibinfo{author}{Press, D.} \emph{et~al.}
\newblock \bibinfo{title}{Ultrafast optical spin echo in a single quantum dot}.
\newblock \emph{\bibinfo{journal}{Nat. Photonics}}
  \textbf{\bibinfo{volume}{4}}, \bibinfo{pages}{367--370}
  (\bibinfo{year}{2010}).

\bibitem{Vezvaee2023PRX}
\bibinfo{author}{Vezvaee, A.}, \bibinfo{author}{Takou, E.},
  \bibinfo{author}{Hilaire, P.}, \bibinfo{author}{Doty, M.~F.} \&
  \bibinfo{author}{Economou, S.~E.}
\newblock \bibinfo{title}{Avoiding leakage and errors caused by unwanted
  transitions in lambda systems}.
\newblock \emph{\bibinfo{journal}{PRX Quantum}} \textbf{\bibinfo{volume}{4}},
  \bibinfo{pages}{030312} (\bibinfo{year}{2023}).

\bibitem{Bodey2019npj}
\bibinfo{author}{Bodey, J.} \emph{et~al.}
\newblock \bibinfo{title}{Optical spin locking of a solid-state qubit}.
\newblock \emph{\bibinfo{journal}{npj Quantum Inf.}}
  \textbf{\bibinfo{volume}{5}}, \bibinfo{pages}{95} (\bibinfo{year}{2019}).

\bibitem{Takou2021PRB}
\bibinfo{author}{Takou, E.} \& \bibinfo{author}{Economou, S.~E.}
\newblock \bibinfo{title}{Optical control protocols for high-fidelity spin
  rotations of single $\mathrm{Si}{\mathrm{v}}^{\ensuremath{-}}$ and
  $\mathrm{Sn}{\mathrm{v}}^{\ensuremath{-}}$ centers in diamond}.
\newblock \emph{\bibinfo{journal}{Phys. Rev. B}}
  \textbf{\bibinfo{volume}{104}}, \bibinfo{pages}{115302}
  (\bibinfo{year}{2021}).

\bibitem{Nguyen2019PRB}
\bibinfo{author}{Nguyen, C.~T.} \emph{et~al.}
\newblock \bibinfo{title}{An integrated nanophotonic quantum register based on
  silicon-vacancy spins in diamond}.
\newblock \emph{\bibinfo{journal}{Phys. Rev. B}}
  \textbf{\bibinfo{volume}{100}}, \bibinfo{pages}{165428}
  (\bibinfo{year}{2019}).

\bibitem{Bauch2018PRX}
\bibinfo{author}{Bauch, E.} \emph{et~al.}
\newblock \bibinfo{title}{Ultralong dephasing times in solid-state spin
  ensembles via quantum control}.
\newblock \emph{\bibinfo{journal}{Phys. Rev. X}} \textbf{\bibinfo{volume}{8}},
  \bibinfo{pages}{031025} (\bibinfo{year}{2018}).

\bibitem{Bauch2020PRB}
\bibinfo{author}{Bauch, E.} \emph{et~al.}
\newblock \bibinfo{title}{Decoherence of ensembles of nitrogen-vacancy centers
  in diamond}.
\newblock \emph{\bibinfo{journal}{Phys. Rev. B}}
  \textbf{\bibinfo{volume}{102}}, \bibinfo{pages}{134210}
  (\bibinfo{year}{2020}).

\bibitem{Taylor2008Nature}
\bibinfo{author}{Taylor, J.~M.} \emph{et~al.}
\newblock \bibinfo{title}{High-sensitivity diamond magnetometer with nanoscale
  resolution}.
\newblock \emph{\bibinfo{journal}{Nat. Phys.}} \textbf{\bibinfo{volume}{4}},
  \bibinfo{pages}{810--816} (\bibinfo{year}{2008}).

\bibitem{Fessler2003}
\bibinfo{author}{Fessler, J.~A.} \& \bibinfo{author}{Sutton, B.~P.}
\newblock \bibinfo{title}{Nonuniform fast fourier transforms using min-max
  interpolation}.
\newblock \emph{\bibinfo{journal}{IEEE Trans. Signal Process.}}
  \textbf{\bibinfo{volume}{51}}, \bibinfo{pages}{560--574}
  (\bibinfo{year}{2003}).

\bibitem{Wernecke1977}
\bibinfo{author}{D'Addario, L.} \& \bibinfo{author}{Wernecke, S.}
\newblock \bibinfo{title}{Maximum entropy image reconstruction}.
\newblock \emph{\bibinfo{journal}{IEEE Transactions on Computers}}
  \textbf{\bibinfo{volume}{C-26}}, \bibinfo{pages}{351--364}
  (\bibinfo{year}{1977}).

\bibitem{Kazimierczuk2011}
\bibinfo{author}{Kazimierczuk, K.}, \bibinfo{author}{Misiak, M.},
  \bibinfo{author}{Stanek, J.}, \bibinfo{author}{Zawadzka-Kazimierczuk, A.} \&
  \bibinfo{author}{Ko{\'{z}}mi{\'{n}}ski, W.}
\newblock \emph{\bibinfo{title}{Generalized Fourier Transform for Non-Uniform
  Sampled Data}}, \bibinfo{pages}{79--124} (\bibinfo{publisher}{Springer Berlin
  Heidelberg}, \bibinfo{address}{Berlin, Heidelberg}, \bibinfo{year}{2012}).

\bibitem{Holland2011}
\bibinfo{author}{Holland, D.}, \bibinfo{author}{Bostock, M.},
  \bibinfo{author}{Gladden, L.} \& \bibinfo{author}{Nietlispach, D.}
\newblock \bibinfo{title}{Fast multidimensional nmr spectroscopy using
  compressed sensing}.
\newblock \emph{\bibinfo{journal}{Angew. Chem. Int. Ed.}}
  \textbf{\bibinfo{volume}{50}}, \bibinfo{pages}{6548--6551}
  (\bibinfo{year}{2011}).

\bibitem{Jaravine2006}
\bibinfo{author}{Jaravine, V.}, \bibinfo{author}{Ibragimov, I.} \&
  \bibinfo{author}{Orekhov, V.}
\newblock \bibinfo{title}{Removal of a time barrier for high-resolution
  multidimensional nmr spectroscopy}.
\newblock \emph{\bibinfo{journal}{Nat. Methods}} \textbf{\bibinfo{volume}{3}},
  \bibinfo{pages}{605--607} (\bibinfo{year}{2006}).

\bibitem{paladino20141}
\bibinfo{author}{Paladino, E.}, \bibinfo{author}{Galperin, Y.~M.},
  \bibinfo{author}{Falci, G.} \& \bibinfo{author}{Altshuler, B.~L.}
\newblock \bibinfo{title}{1/f noise: Implications for solid-state quantum
  information}.
\newblock \emph{\bibinfo{journal}{Rev. Mod. Phys.}}
  \textbf{\bibinfo{volume}{86}}, \bibinfo{pages}{361--418}
  (\bibinfo{year}{2014}).

\bibitem{Medford2012PRL}
\bibinfo{author}{Medford, J.} \emph{et~al.}
\newblock \bibinfo{title}{Scaling of dynamical decoupling for spin qubits}.
\newblock \emph{\bibinfo{journal}{Phys. Rev. Lett.}}
  \textbf{\bibinfo{volume}{108}}, \bibinfo{pages}{086802}
  (\bibinfo{year}{2012}).

\bibitem{Bruzewicz2019APR}
\bibinfo{author}{Bruzewicz, C.~D.}, \bibinfo{author}{Chiaverini, J.},
  \bibinfo{author}{McConnell, R.} \& \bibinfo{author}{Sage, J.~M.}
\newblock \bibinfo{title}{Trapped-ion quantum computing: Progress and
  challenges}.
\newblock \emph{\bibinfo{journal}{Appl. Phys. Rev.}}
  \textbf{\bibinfo{volume}{6}}, \bibinfo{pages}{021314} (\bibinfo{year}{2019}).

\bibitem{Ruster2016APB}
\bibinfo{author}{Ruster, T.} \emph{et~al.}
\newblock \bibinfo{title}{A long-lived zeeman trapped-ion qubit}.
\newblock \emph{\bibinfo{journal}{Appl. Phys. B}}
  \textbf{\bibinfo{volume}{122}}, \bibinfo{pages}{254} (\bibinfo{year}{2016}).

\end{thebibliography}
\end{document}